\documentclass{sig-alternate-10pt}
\pdfoutput=1

\makeatletter

\providecommand{\tabularnewline}{\\}

\newcommand{\comment}[1]{}
\usepackage{algorithm}
\usepackage[noend]{algorithmic}
\usepackage{url}

\begin{document}

\title{Nation-State Routing: Censorship, Wiretapping, and BGP}

\author{Josh Karlin, Stephanie Forrest, and Jennifer Rexford}

\maketitle
\begin{abstract}
  The treatment of Internet traffic is increasingly affected by
  national policies that require the ISPs in a country to adopt common
  protocols or practices.  Examples include government enforced
  censorship, wiretapping, and protocol deployment mandates for IPv6
  and DNSSEC. If an entire nation's worth of ISPs apply common
  policies to Internet traffic, the global implications could be
  significant.  For instance, how many countries rely on China or
  Great Britain (known traffic censors) to transit their traffic?
  These kinds of questions are surprisingly difficult to answer, as
  they require combining information collected at the prefix,
  Autonomous System, and country level, and grappling with incomplete
  knowledge about the AS-level topology and routing policies.  In this
  paper we develop the first framework for country-level routing
  analysis, which allows us to answer questions about the influence of
  each country on the flow of international traffic.  Our results show
  that some countries known for their national policies, such as Iran
  and China, have relatively little effect on interdomain routing,
  while three countries (the United States, Great Britain, and
  Germany) are central to international reachability, and their
  policies thus have huge potential impact.
\end{abstract}

\section{Introduction}

Internet routing is typically studied at the Autonomous System (AS)
level. This is by design.  Traditionally, ASes control their own
internal networks and set their own policies for the routing,
filtering, and monitoring of traffic, placing policy in the hands of
the organizations that own them.  Recently, groups of ASes have begun
to act under common policies, issued by their country's
government. Examples include Internet censorship~\cite{ONIBOOK},
wiretapping~\cite{PAA}, and protocol-deployment
mandates~\cite{gov:dns,gov:ipv6}. For instance, Chinese, British, and
Pakistani ISPs are required (or strongly encouraged) to filter content
deemed socially offensive. Although censoring techniques differ, all
three countries are known to block traffic at the IP level (e.g., by
filtering based on IP addresses and URLs in the data packets, or
performing internal prefix
hijacks~\cite{Clayton05,youtube,Crandall07}), which could affect the
international traffic they transit.  Some countries, such as the
United States and Sweden, wiretap international traffic, where even
encrypted traffic is vulnerable to traffic-analysis
attacks~\cite{XY05}. Finally, governments can attempt to force the
deployment of protocols, such as the deployment of IPv6 and DNSSEC in
federal agencies of the United States.

It is unclear what effect any particular country's policies have on
the rest of the Internet. Typically, censorship is applied to prevent
domestic users from reaching disagreeable content.  However, some
censorship techniques (such as filtering based on IP
addresses or URLs) may affect all traffic traversing an AS.  In
addition, ASes might intentionally, or accidentally as in the recent
YouTube outage~\cite{youtube}, apply censorship policies to
international traffic.  How many networks outside of the country would
be prevented from viewing Web pages simply because their traffic
traverses one of these networks?  Which international traffic is
vulnerable to warrantless wiretapping by the United States or Sweden?
And, finally, how feasible is it to avoid directing traffic through a
given country with objectionable policies by using alternative routes?

To answer these questions, we must study the aggregate effect of
national policies on the flow of international traffic, rather than
analyzing individual ASes in isolation.
In this paper we take initial steps toward understanding interdomain
routing at the nation-state level.  We are particularly interested in
understanding the influence that each country's ASes have over
reachability between other countries. The resulting data and
measurement techniques could be useful to many communities. First,
those regions of the world with strong dependencies on particular
countries could use our result to guide changes in how they connect to
the rest of the Internet.  Second, overlay networks (such as Resilient
Overlay Networks~\cite{RON}) could use our results to determine how
best to circumvent specific countries, helping to ensure that data are
delivered intact, and avoid snooping.  Third, our results would be
helpful to policy makers to understand what impact their decisions
could have on the global Internet.

There are two primary challenges in this work. The first is to define
suitable metrics for quantifying the importance, or centrality, of
each country to Internet reachability. The second is to accurately
infer the data needed to compute these metrics, and validate them. We
adapt the {\em betweenness centrality} metric from statistical physics
as a first approximation of country centrality.  Betweenness
centrality is typically used as a naive traffic estimator at each node
in a graph.  We adapt betweenness centrality to estimate the impact
each country has on reachability between other countries, defining
{country centrality} (CC) in Section~\ref{sec:Reachability-Metrics}.

Our metrics take as input the country-level paths between each pair of
IP addresses in the Internet. This is a significant challenge because
of the many levels of inference required to produce a country-level
interdomain path. First, ASes select routes using the Border Gateway
Protocol (BGP)~\cite{Rekhter06}, which chooses routes based on
undisclosed routing policies, rather than simply using the shortest
path. Fortunately, this is a well-studied problem and several
inference algorithms exist for inferring AS-level routes.  A second
challenge arises because an individual AS may span many countries.
This leads us to consider routing at the IP prefix level, which
requires understanding how packets traverse each AS. Finally, each
path must be converted to a country-level path by mapping IP addresses
to prefixes, and then prefixes to countries (e.g., using routing
registry data).  There is a risk of introducing significant, and
possibly compounding, error in each step of the process. However, we
present empirical evidence to suggest that our centrality metric is
robust to the measurement noise, and that our results are meaningful.

Our inference techniques allow us to estimate the centrality of each
country, where CC values range from $0$ (implying no influence) to $1$
(the theoretical maximum).  Our results show that countries known for
censorship, such as Great Britain, China, Australia, and Iran, have CC
values of 0.29, 0.07, 0.07, and 1.12e-05 respectively. These results
suggest that, of the countries that censor Internet traffic, only some
have significant impact on global routing.  In particular, the
countries that have received the most publicity for their censorship,
such as China, have significantly less impact on international traffic
than, say, Great Britain, which also censors traffic. We also show
that the United States and Sweden (nations known to permit warrantless
wiretapping) have CC values of 0.74 and 0.02; even if ASes actively
prefer BGP routes that avoid the United States, the CC value only
drops from 0.74 to 0.55.

With national policies on the rise, we believe that researchers, ISPs,
and policy makers will soon need to understand the impact that these
policies can have on other countries, networks, and even individual IP
prefixes. Our major contribution is the development of a framework for
studying interdomain routing at the nation-state level. This includes
identifying and addressing the many challenges of inferring the
country-level paths, developing network centrality metrics appropriate
for the problem, validating the methods, and reporting initial results.

The paper is organized as follows. In the next section we briefly
discuss the Internet's topology and the correct granularity for
measuring country paths. In Section~\ref{sec:Populating-the-Model} we
design, implement, and validate the {\em Country Path Algorithm} (CPA)
for inferring country-paths from a pair of source and destination IP
addresses. The algorithm has several stages, as it must first infer
the interdomain path, and then intradomain paths, and finally
determine the country path.  Next,
Section~\ref{sec:Reachability-Metrics} reviews betweenness centrality
and presents two extensions for measuring a country's influence over
global reachability. These metrics take as input the global
measurements produced by the CPA.  In Section~\ref{sec:Results}, we
apply our inference techniques to sample data sets of traceroutes and
AS paths, as well as inferred paths between all known IP
prefixes. This helps validate that our metrics are robust to inference
error.  We also present initial results characterizing the data
produced by the CPA. Next, we discuss future work and other possible
challenges in country level analysis in Section~\ref{sec:Discussion},
we review related work in Section~\ref{sec:Related-Work}, and
finally conclude in Section~\ref{sec:Conclusion}.

\section{An Appropriate Granularity for Analyzing Country-Level Paths}
\label{sec:Background}

\begin{figure}
\includegraphics[width=0.95\columnwidth]{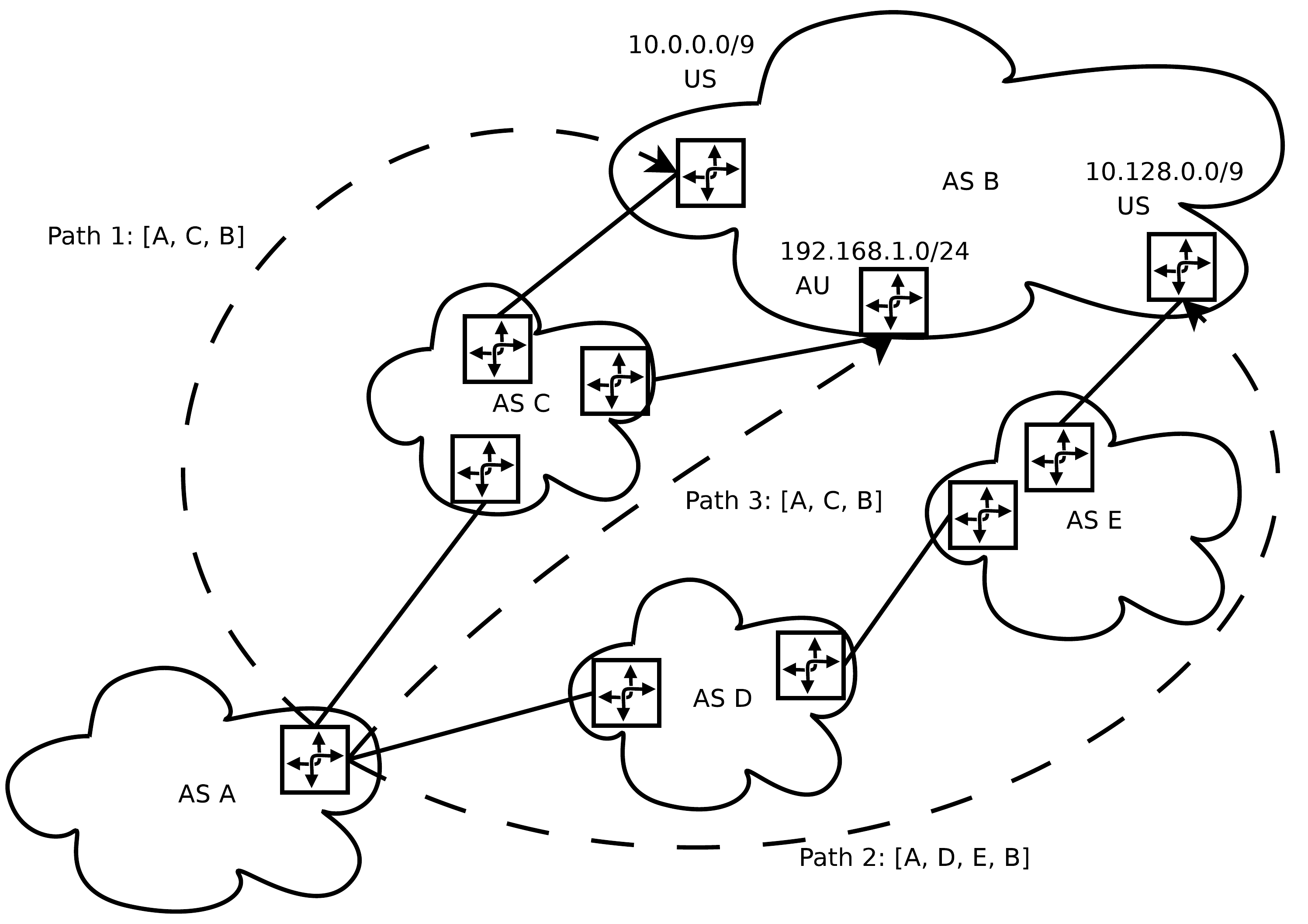}
\caption{Example AS topology with AS paths. Paths 1 and 2 both route
between the same pair of ASes (A and B), but their AS paths are
different, depending on the destination prefix. The same AS path can
also have distinct country-level paths, for example paths 1 and 3. }
\label{fig:multipaths}
\end{figure}

The Internet is currently comprised of roughly 30,000 Autonomous
Systems, which are typically independently operated, multi-homed
networks.  Each AS is allocated IP address space, which is a contiguous
blocks of IP addresses called IP prefixes. The interdomain routing
protocol that allows ASes to reach one another's prefixes is called
the Border Gateway Protocol (BGP). BGP is a policy based protocol that
selects and propagates AS paths according to local policies (e.g.,
economic relationships), rather than path performance (e.g., shortest
path routing). Example policies include customer-provider in which the
customer pays the provider for transit, and peer-peer in which the
participating ASes transit each other's customer traffic to their own
customers. Lixin Gao showed that these policies can affect routing
propagation ~\cite{gao01}. For instance, most customers would not be
willing to provide transit from one of their providers to another.
Gao observed that ASes typically follow the {\em valley-free} rule,
which states that routes received from a provider or peer should only
be propagated to customers.

For our experiments it is necessary to inference all of the
country-paths between each pair of IP addresses.  Since IP addresses
are allocated to ASes, we could determine the country-paths between
each pair of ASes and use that information to determine all paths
between each pair of IP addresses.  One immediate problem is that some
ASes span more than a single country.  A second issue is that in many
cases there are multiple paths between two ASes, depending on where
traffic enters the AS and on the destination prefix in question. For
example, in Figure \ref{fig:multipaths} AS A uses path 1 to reach
prefix 1 at AS B, but uses path 2 to reach prefix 2 at the same
destination AS. AS B might split its traffic like this to balance its
traffic load between two providers (ASes C and E). 

A second possible approach would cluster together the prefixes with
the same AS paths between AS pairs, and infer a path for one prefix
from each cluster.  This is known as a {\em BGP Atom} ~\cite{Broido01,Broido01-2}.
Although this approach can enumerate the best AS-paths between AS
pairs, it does not encompass the full diversity of country-level
paths. Two destination prefixes with the same AS path may have
different underlying country-level paths. For instance, in Figure
\ref{fig:multipaths} AS paths 1 and 3 are the same, however they
terminate in different countries (United States in path 1 Australia in
path 3).

After ruling out the first two approaches, we resorted to inferring
the country-level paths between each pair of IP prefixes, the finest
level of measurement available. There are over 290,000 prefixes in
today's routers, resulting in over 84 billion country paths that need
to be inferred and analyzed. We also study all of the available
alternate paths that exist from one prefix to another, resulting in
more than 465 billion country path inferences that need to be
performed.  The large number of inferences places significant
constraints on the inference algorithm's complexity.  For instance,
simply running Dijkstra's shortest path algorithm to determine the
intradomain path of each AS in each path is too slow.

\section{The Country Path Algorithm}
\label{sec:Populating-the-Model}

\begin{figure}
\[
traceroute=\overbrace{ip_{src},ip_{2},ip_{3}}^{C_{1}},\overbrace{ip_{4},ip_{5},ip_{6}}^{C_{2}},\overbrace{ip_{dst}}^{C_{3}}\]

\[
traceroute=\underbrace{ip_{src},ip_{2}}_{AS_{1}}\underbrace{,ip_{3},ip_{4},ip_{5},}_{AS_{2}}\underbrace{ip_{6},ip_{dst}}_{AS_{3}}\]

\begin{centering}
Country-path Inference Algorithm: $(ip_{src},ip_{dst})\rightarrow(AS_{1},AS_{2},AS_{3})\rightarrow(C_{1},C_{2},C_{3})$
\par\end{centering}
\caption{Traceroutes, AS-paths, and country-paths.  A traceroute is
  the list of IP addresses of the routers that a packet traverses from
  $ip_{src}$ to $ip_{dst}$. Each router belongs to an AS, and each
  router is in a country C. The Country Path Algorithm takes a source and
  destination IP address as input, infers the interdomain AS-path
  between the two addresses, and then infers the country-path between
  them.}

\label{fig:overview}
\end{figure}

The metrics described in Section \ref{sec:Reachability-Metrics}
analyze country-level paths to determine which countries can
potentially interfere with the communication of others.  In this
section we present the Country Path Algorithm (CPA) for inferring the
country-level paths between any two IP addresses.  There are two steps
to the procedure.  The first infers the interdomain path between the
addresses, and the second step predicts the country-path from the
AS-path.  We use a slightly modified version of Qiu et al.'s
~\cite{Qiu05} AS-path heuristic for the first step which is described
in \ref{sub:Prefix-Pair-toAS}, and introduce the first country path
predictor in the second step, presented in
\ref{sub:AS-path-to-Country-path}. An overview of the CPA algorithm is
shown in Figure \ref{fig:overview}.  The AS-path to country-path
heuristic requires information about known traceroutes and their
corresponding AS-paths and country-paths as input. We show how to
infer these paths from a traceroute in Subsection
\ref{sub:IP-Path-to-Country-Path}.

\subsection{Prefix Pair to AS-path}\label{sub:Prefix-Pair-toAS}

The first step in the country path algorithm is to map prefix
source/destination pairs to their appropriate AS paths.  Of the recent
AS-path inference methods
~\cite{Qiu05,Mao05,Muhlbauer06,Madhyastha06}, only Qiu's provides
prefix-level predictions and is fast enough for our needs.

\subsubsection{A Modified Version of Qiu's Heuristic}

\begin{figure}[t]
\begin{centering}
{\footnotesize 
\begin{algorithmic}[1]
\STATE KnownPath(p, G, prePaths):
\WHILE{queue.length $>$ 0} 
\STATE u $\leftarrow$ POP(queue,0)
\FORALL{v $\in$ peers(u)}
\STATE $P_u \leftarrow$ ribIn(u)[p][0]
\IF{legitimatePath((v)+$P_u)$}
\STATE tmppath $\leftarrow$ ribIn(v)[p][0]
\STATE update ribIn(v)[p] $\leftarrow$ with (v) + $P_u$ \label{update}
\STATE sort(ribIn(v)[p])
\IF{tmppath = path(v)[p][0] and v $\in$ queue}
\STATE append(queue,v)
\ENDIF
\ENDIF
\ENDFOR
\ENDWHILE
\RETURN ribIn
\end{algorithmic}
}
\par\end{centering}{\footnotesize \par}

\caption{Pseudo-code of Qiu's inference algorithm. Line 6 was modified 
to propagate paths to pre-determined ASes.}
\label{fig:KnownPath}
\end{figure}
\begin{figure}[t]
\begin{centering}
{\footnotesize 
\begin{algorithmic}[1]
\STATE ComparePath($P_1 = (u,v1,...), P_2 = (u,v2,...)$):
\IF{$P_1$.ulen $\neq$ $P_2$.ulen}
\RETURN $P_1$.ulen - $P_2$.ulen
\ENDIF
\IF{$|P_1| \neq |P_2|$}
\RETURN $|P_1| - |P_2|$
\ENDIF
\IF{$P_1$.freq $\neq$ $P_2$.freq}
\RETURN $P_2$.freq - $P_2$.freq
\ENDIF
\RETURN $P_1-P_2$
\end{algorithmic}
}
\par\end{centering}{\footnotesize \par}

\caption{Pseudo-code of Qiu's path comparison heuristic. Lines 2-4 have been
switched with lines 5-7 from the original algorithm.}
\label{fig:ComparePathSPF}
\end{figure}

Qiu's heuristic simulates the propagation of BGP routes across an
AS topology, as if each AS had a single router. The propagation model
is a simplified model of the actual BGP protocol. In it, each router
selects its best path to the destination prefix after receiving a route
announcement, and propagates the path to its neighbors (obeying the
valley-free rule) if its best path has changed. The largest contribution
that her work made was to include known BGP paths from routing table
dumps (known as RIBs) to improve the accuracy of the heuristic. Essentially,
ASes are primed with known paths for each prefix at the beginning
of the algorithm. Then, as the paths are propagated, paths that are
the fewest hops from a known path are given preference.

The original Qiu algorithm does not propagate paths to ASes that have
pre-determined paths, since they will never select an alternate path.
Therefore, many ASes will only have a best path, and no selection of
alternative paths available. Our centrality metrics require a list of
all possible alternate paths for each AS to each prefix as well as the
best path. This is needed to estimate the ability of networks to route
around (or avoid) particular countries using alternate paths.
Therefore, we modified Qiu's algorithm to propagate paths to all ASes,
even those that were primed with a known path. Our changes to the
original algorithm, are shown in Figures \ref{fig:KnownPath} and
\ref{fig:ComparePathSPF}. The purpose of our alterations is to predict
alternate paths, not to increase the algorithm's accuracy. In the
validation section we show that our changes appear to have no
significant effect on the predictive accuracy of the algorithm.

\subsubsection{Pre-processing the Data}

Qiu's algorithm takes as input a list of known BGP routes and a topology
of known ASes, the edges between ASes, and the economic relationship of each
edge. We retrieved the first RIB of 2009 (BGP routing table) from
RouteViews ~\cite{oregon05} and RIPE RIS servers ~\cite{ripe}. In total there
are paths for 290,691 prefixes. We divided the data in half, into
a testing and training set. All routes from each observation point
are kept together, and all observation points in the same AS are
also kept together. 

The topology was extracted from the AS paths found in the BGP RIBs.
We developed a topology for use in testing and the total set for use
in our final experiments. The training set topology has 29,580 vertices
(ASes) and 68,396 edges while the total set has 29,607 vertices and
77,683 edges. 

The edges of the topology must be labeled as one of customer-provider,
peer-peer, or sibling-sibling (two AS numbers that represent the same
network). We implemented the relationship inference algorithm described
in ~\cite{gao01} and labeled the edges of our topologies with the results.
In total, the testing topology has 6,616 peer-peer edges, 61,037 customer-provider
edges, and 743 sibling-sibling edges. The total topology has 12,623
peer-peer edges, 64,050 customer-provider edges, and 1,010 sibling-sibling
edges.

\subsubsection{Validation}

To ensure that our implementation of the heuristic was working
correctly, we downloaded RouteViews and RIPE RIBS from the beginning
of 2005, which is close in time to the data used for Qiu et
al.'s\emph{ }original paper. We split the data into testing and
training sets proportional in size to the data sets used in
~\cite{Qiu05} (we used the RIPE data for training, and tested on the
RouteViews data), and then fed the testing topology and paths as input
to the heuristic for prediction of paths in the testing set. The
heuristic was able to predict 60\% of the testing paths, exactly as
stated in the original paper. This shows that the alterations had
little effect on the algorithm, and suggests that our implementation
is correct.

\comment{\footnote{Due to some last minute changes to the algorithm,
  we were unable to run the exact algorithm shown here for all of our
  AS path inference based results in the paper. Our preliminary
  results show that the changes have little impact on the results
  (roughly 1\% prediction accuracy difference), and the final
  submission will include updated results with the exact algorithm
  shown here.}}

On our 2009 data set, the algorithm is able to predict the exact path
found in the training set of the RIB correctly 54\% of the
time. However, the exact path is often in the routing table, but not
selected as the best path. We show that the exact path is in the
routing table 80\% of the time.

Our results suggest that the routing table of each AS is relatively
accurate, however the best path is not reliably selected. We return
to this point in Section \ref{sec:Reachability-Metrics} and show
experimentally that the heuristic is accurate enough for the reachability
analysis that we perform.

\subsection{Mapping Traceroutes to AS and Country Paths}\label{sub:IP-Path-to-Country-Path}

The next step is to map an AS-path into a country-path. This requires
information about known country-level paths and their respective AS-paths.
In this sub-section we describe how we extract country-level and AS-level
paths from traceroutes, and the next section shows how the data can
be used for inference country-level paths.

\subsubsection{Challenges}

Traceroutes show the router-level path between two IP addresses. By
converting the routers' IP addresses to countries, we can determine
the countries that a packet traverses. 

There are many impediments to this process. First, a router can mask
its existence in traceroutes by not decrementing packet TTLs, but
we assume that this is a rare practice. A router could also be configured
to not respond to traceroutes, which happens relatively frequently. Such traceroutes
are incomplete, but we can still extract useful information from them.

The next challenge is to understand the location (country and AS) of
each IP addresses found in the traceroutes. IP addresses are allocated 
to ASes by the regional routing registries (ARIN, RIPE, AFRINIC,
APNIC, and LACNIC). Each regional registry publishes a database of
allocated IP space, the ASes they were allocated, and the country of the
organization. Once allocated, it is up to the ASes to update the
registry databases of any changes. For instance, if an ISP delegates a
portion of its prefix to a customer AS, that customer should be
registered for the particular sub-prefix. This is not always done, and
the registries are known to be incomplete and often inaccurate
~\cite{mahajan02,Sriram09}.

\subsubsection{Algorithm and Data}
We collected traceroutes from the iPlane project ~\cite{iplane} on December
17th of 2008. The data set contains roughly 26 million traceroutes,
that were collected from 198 observation points (the majority of which
are PlanetLab ~\cite{planetlab} nodes), with an average of 133,580 traceroutes
each. 

To convert the traceroutes to country-paths, we first had to obtain
registry information for each IP address in the traceroutes. Team
Cymru ~\cite{cymru} keeps track of registry allocated prefixes and associated
country code and AS mappings. For each IP in the traceroutes (as well
as each prefix in the RIBs), we queried Team Cymru's server to obtain
the country code. In the case that the lookup failed, or that the
response was vague, such as {}``EU'' (Europe) or {}``AP'' (Asia
Pacific), we ran a normal whois request (version 4.7.27) and extracted
country and AS information where possible (whois responses vary,
some contain more information than others). Our only tweak to the
data was to replace the Hong Kong country code with China since they
are now the same country. In total, we were able to determine a specific
country code for 99\% of the IPs found in traceroutes.

\subsubsection{Validation}

To verify the accuracy of our IP to country code and AS lookups, we
compared our results to known ASes and countries for particular
routers. One method of extracting the actual location of a given
router is to extract it from its DNS hostname. For instance, the
router with hostname, 143.ATM3-0.XR2.LAX2.ALTER.NET, is located in Los
Angeles, which is in the United States. Two projects have developed
hostname to location heuristics, RocketFuel's undns ~\cite{rocketfuel}
and the sarangworld project ~\cite{sarangworld}, and the iPlane
project has applied them to the routers in the traceroute data set. The
locations were further verified by the iPlane project by timing
analysis and known topology information. 

For each IP address that was resolved to a country and AS using undns
and sarangworld (9\% of IPs in the traceroutes), we compared the values
to our inferenced data from routing registries. We found that we could
correctly infer the country of a router 96\% of the time, and the
AS 92\% of the time. Our verification suggests that we have relatively
accurate data sets with which to build our AS Path to country-path
heuristic.

\subsection{AS-path to Country-path}\label{sub:AS-path-to-Country-path}

The last piece of our IP address pair to country-path algorithm involves
inferring a country-path from an AS path. In total, the final algorithm
takes a pair of IP addresses as input, determines their longest matching
prefixes (like a routing table lookup), finds the best inferred AS
path between them, and finally uses the algorithm in this sub-section
to infer the countries along the path.

\subsubsection{Challenges}
It is difficult to infer intradomain routes.  An Autonomous System is
so called because it has complete control over its intradomain
network. It can use whatever protocols it likes, even experimental
ones, with its own policies, to determine how packets traverse its own
network. This makes it very difficult for an outsider to determine how
a packet might route through an AS.  We do know that common
intradomain protocols (e.g. OSPF~\cite{OSPF} and IS-IS~\cite{ISIS})
will typically choose the shortest path between any two points in the
network. The difficulty is that the definition of shortest path can
change between networks. For some networks, a short path might be low
latency, where for others it might be one that follows a
high-bandwidth path.

Since we are provided with an inferred AS path, the next step is to
determine where the route will enter (ingress router) and exit (egress
router) each AS.  A simply heuristic for finding the exit router might
be to find the the nearest router to the ingress router that is
connected to the next hop AS.  But again, nearness is not well
defined.

Finally, the algorithm has to be fast enough to infer a country-path
for 465 billion interdomain paths (one inference for each pair of
prefixes over each AS path in each router's RIB) in a reasonable
amount of time. Performing Dijkstra's shortest path across large ASes
with tens of thousands of routers billions of times is simply too
slow, and most AS paths include at least one AS of that size.

\subsubsection{The Algorithm}

\begin{figure}
\[
\underbrace{ip_{src},ip_{2}}_{AS_{1}}\underbrace{,ip_{3},ip_{4},ip_{5},}_{AS_{2}}\underbrace{ip_{6},ip_{dst}}_{AS_{3}}\]

\caption{Example annotated traceroute. $ip_{src}$, $ip_{3}$,
and $ip_{6}$ are AS ingress points, and $ip_{2}$ and $ip_{5}$ are
AS egress points. }

\label{fig:triple}
\end{figure}

\comment{
\begin{figure}[t]
\begin{centering}
{\footnotesize 
\begin{algorithmic}[1]
\STATE Initialize(Traceroutes):

\FOR{route in Traceroutes}
\STATE $AS$ = ASs in route
\STATE $IP$ = ingress point for each AS
\STATE $C$ = countries between each ingress point
\FOR{$(i = 0;\; i < |AS|-1;\; i++)$}
\IF{$IP_i$ or $IP_{i+1}$ = no traceroute response}
\STATE \textbf{continue}
\ENDIF
\STATE $freq_s$[$AS_{i+1}$][$IP_{i+1}$] += 1
\STATE $freq_{sc}$[$AS_{i+1}$][$C_i[0]$][$IP_{i+1}$] += 1
\STATE $freq_d$[$AS_{i}$][$AS_{i+1}$][$IP_{i+1}$] = (+1,$C_i$)
\STATE $freq_{dc}$[$AS_{i}$][$AS_{i+1}$][$C_i[0]$][$IP_{i+1}$] = (+1,$C_i$)
\STATE $known_s$[$AS_i$][$AS_{i+1}$][$IP_i$] = ($IP_{i+1}$,$C_i$)
\IF{$(i < |AS|-2)$}
\STATE $known_d$[$AS_i$][$AS_{i+1}$][$AS_{i+2}$][$IP_i$] = ($IP_{i+1}$,$C_i$)
\ENDIF
\ENDFOR
\ENDFOR
\STATE $maxFreq_s$ = max. ingress $IP$ count for each $AS$
\STATE $maxFreq_{sc}$ = '' '' ($AS,C$) pair
\STATE $maxFreq_d$ = max. ($IP$,$C$) for each ($AS,AS_{i+1}$) pair
\STATE $maxFreq_{dc}$ = max. ($IP$,$C$) for each ($AS,AS_{i+1},C$) triple
\end{algorithmic}
}
\par\end{centering}{\footnotesize \par}

\caption{Pseudo-code of AS-path to country-path initialization function.}
\label{fig:Initialize}
\end{figure}

\begin{figure}[t]
\begin{centering}
{\footnotesize 
\begin{algorithmic}[1]
\STATE PredictCountryPath($AS$,$fromIP$,$toPrefix$):
\STATE $Path_C$ = empty list
\STATE $fromPrefix$ = longest matching prefix for FromIP
\STATE $entry$ = $fromIP$
\FOR {$(i = 0;\; i < |AS|-1;\; i++)$}
\STATE $cs$ = empty list
\STATE $curC$ = tail($Path_C$)
\STATE $nextIP$ = unknown
\IF {exists($known_d$[$AS_i$][$AS_{i+1}$][$AS_{i+1}$][$entry$])}
\STATE ($nextIP,cs$) = $known_d$[$AS_i$][$AS_{i+1}$][$AS_{i+1}$][$entry$]
\ELSIF{exists($known_s$[$AS_i$][$AS_i+1$])}
\IF{exists($known_s$[$AS_i$][$AS_{i+1}$][$entry$])}
\STATE ($nextIP,cs$) = $known_s$[$AS_i$][$AS_{i+1}$][$entry$]
\ELSE
\IF {exists($maxFreq_{dc}$[$AS_i$][$AS_{i+1}$][$curC$])}
\STATE ($nextIP,cs$) = $maxFreq_{dc}$[$AS_i$][$AS_{i+1}$][$curC$]
\ELSE
\STATE ($nextIP,cs$) = $maxFreq_{d}$[$AS_i$][$AS_{i+1}$]
\ENDIF
\ENDIF
\ELSE
\IF {exists($maxFreq_{sc}$[$AS_{i+1}$][$curC$])}
\STATE $nextIP$ = $maxFreq_{sc}$[$AS_{i+1}$][$curC$]
\ELSE
\STATE $nextIP$ = $maxFreq_s$[$AS_{i+1}$]
\ENDIF
\ENDIF
\STATE $entry$ = $nextIP$
\STATE append($Path_C$, $cs$)
\ENDFOR
\STATE append($Path_C$, country of toPrefix)
\STATE $Path_C$ = remove consecutive duplicates
\RETURN $Path_C$

\end{algorithmic}
}

\par\end{centering}{\footnotesize \par}

\caption{Pseudo-code of AS-path to country-path prediction function.}
\label{fig:Prediction}
\end{figure}

} 

\begin{figure}[t]
\begin{centering}
{\footnotesize
\begin{algorithmic}[1]
\STATE predictCountries(AS-path):
\STATE 
\FOR{each ASN in the AS-path}
\IF {(a known ingress point exists for the next ASN from this ingress)}
\STATE Select countries and next ingress point from known-ingress
\ELSIF{(a known ingress point exists for the next ASN from this ASN in this country)}
\STATE Select most frequented ingress point (and corresponding country path)
\ELSIF{(a known ingress point exists for the next ASN from this ASN)}
\STATE ``''
\ELSIF{(a known ingress point exists for the next ASN from this country)}
\STATE ``''
\ELSIF{(a known ingress point exists for the next ASN)}
\STATE ``''
\ENDIF
\ENDFOR
\end{algorithmic}
}
\par\end{centering}{\footnotesize \par}
\caption{Pseudo-code of AS-path to country-path prediction}
\label{fig:Prediction}
\end{figure}

We present a linear time (relative to the size of the AS path) algorithm
to inference country-paths from AS-paths. The insight of the algorithm,
similar to Qiu's AS-path algorithm, is to use known intradomain paths
as often as possible, rather than infer our own. 

The algorithm is broken down into two phases, initialization, and path
inference. In the initialization phase, the (traceroute, country-path,
AS-path) triples of known data are parsed for two particular features.
First, each AS's ingress point is stored, relative to the ingress
point of the previous AS in the path. For instance, Figure
\ref{fig:triple} shows an example triple in which we learn that when
$AS_{2}$ is entered at $ip_{3}$, and $AS_{3}$ is the next AS, with
ingress point $ip_{6}$. Therefore, when AS path $AS_{2},AS_{3}$ is
seen in the future, and $AS_{2}$ was entered at $ip_{3}$, then we
infer that $ip_{6}$ is $AS_{3}$'s ingress point and will have the
country-path inferred from ip addresses $ip_{3},ip_{4},ip_{5},$and
$ip_{6}$. To increase accuracy, we also look two ASes ahead to
determine the next AS's ingress point. For instance, we learn that
when $AS_{1}$ is entered at $ip_{src}$ and $AS_{2}$ and $AS_{3}$ are
next, then $ip_{3}$ is the ingress point to $AS_{2}$.  We store this
information in a hash table referred to as the known-ingress table.

We will not have a value in the known-ingress table for every
combination of ASes and ingress points. Therefore, it is sometimes
necessary to to guess ingress points for the next AS. To aid in our
guesses, the initialization algorithm also keeps track of the
frequency of each AS's ingress points.  For instance, we might learn
that $ip_{3}$ is the ingress point for $AS_{2}$ 75\% of the time, or
50\% of the time when coming from an AS in Canada, or 90\% of the time
when coming from anywhere in $AS_{1}$. We keep track all of these
frequencies, and their relationships to previous ASes and countries.


The prediction algorithm is shown in Figure \ref{fig:Prediction}.
For each AS in the AS path, it searches the known data for the current
context (e.g. next AS, current country, current ingress point), progressively
becoming less specific, until a match is found. A match provides information
about the next ingress point and the list of countries between the current
and next ingress points. This proceeds until the final ingress point is
found. At which point, the country of the destination prefix is appended
to the country-path and the path is returned.

\subsubsection{Validation}

To validate our algorithm, we selected roughly 1.4 million complete
traceroutes from the testing set in which every router along the path
has been determined the country and AS are known for each router,
and the source and destination IP addresses are from different countries.
Then, we initialized the prediction algorithm with the training set
and predicted country paths for the test routes. Our algorithm predicted
the exact set of countries 78\% of the time. Another way of comparing
the agreement of the predicted results to the known set of paths is
to take the intersection of the sets over the union $\frac{Predicted\cap Actual}{Predicted\cup Actual}$
, as seen in ~\cite{Madhyastha06}. The agreement
between our predicted paths and the actual paths is 92\%, suggesting
that when the predictor is wrong, it is usually close.

\section{Reachability Metrics}\label{sec:Reachability-Metrics}

\begin{figure}
\begin{centering}
\includegraphics[width=0.9\columnwidth]{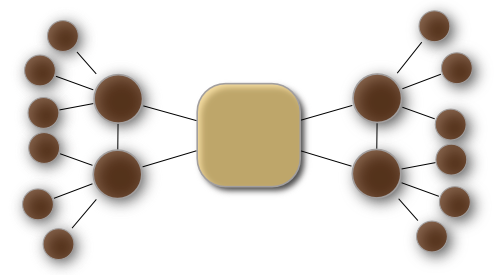}
\par\end{centering}

\caption{Betweenness centrality. The middle node does not have the greatest
degree, but it is along the greatest number of shortest paths.}
\label{fig:betweenness}
\end{figure}

There are many ways to quantify the importance (or centrality) of a
node in a network.  Network centrality is a well studied problem
~\cite{Freeman77,Freeman79,Sabidussi66} in statistical physics that
has recently been applied to the AS-level Internet
~\cite{our:rad,Mahadevan06,Zhou04}. In this section we discuss the
betweenness centrality metric, which is a centrality metric that we
adapt for our own experiments.  From betweenness centrality, we derive
two metrics for measuring the centrality of a country at the BGP
level.

\subsection{Background on Betweenness Centrality}

The simplest centrality metrics measure the degree of a node and the
average shortest-path distance from a node to any other in the network.
More advanced metrics, such as betweenness centrality, directly incorporate
the importance of a node to network routing. 

Betweenness centrality is an estimator of the importance of a node
for communication flow in a network. It assumes that traffic flows
equally along the shortest paths between two points, that each node
has unit traffic, and that each node's traffic is uniformly distributed
to the other nodes. It then estimates how much traffic flows through
each node with the following formula: \[
Betweenness(\upsilon)=\sum_{{s\neq\upsilon\neq t\in V\atop s\neq t}}\frac{\sigma_{s,t}(\upsilon)}{\sigma_{s,t}}\]
 where $\sigma_{s,t}$ is the number of shortest paths between $s$
and $t$ and $\sigma_{s,t}(\upsilon)$ is the number of shortest paths
between $s$ and $t$ that transit through $\upsilon$. Nodes that
transit lots of traffic have higher betweenness values than those
that transit little. Figure \ref{fig:betweenness} depicts an example
network in which the middle node has the highest betweenness, even
though four nodes have greater degree.

If each pair of nodes in the network had a single shortest path between
them, then the betweenness centrality of a node could be interpreted
as the number of shortest paths that pass through the node. In a network
like the Internet, there are typically many shortest paths between
two nodes. When multiple shortest paths exist, betweenness centrality
splits the traffic equally among the shortest paths (by dividing it
by $\sigma_{s,t}$). A node's betweenness centrality then represents
the total amount of traffic it transits, given the stated assumptions.

\subsection{Country Centrality}

In this study, we are interested in determining each country's influence
over global reachability. This is not the same as determining how
much traffic a country transits. Although a country might transit
50\% of all Internet traffic, that does not necessarily imply that
50\% of country-pairs rely upon that country to communicate with one
another. But, traffic estimates can still be useful for determining
influence over reachability.

Because we are concerned with global reachability, we assume that all
countries are equally important, and wish to communicate with one
another uniformly. We then want to determine how much influence each
country has over the communication paths. This can be thought of as a
traffic estimation problem in which all countries have unit traffic,
and all countries split that traffic equally to each destination.
Then, to determine influence, we measure how much traffic each node
transits. This is similar to the problem that betweenness
centrality tries to solve.

There are three significant differences between country centrality and
betweenness centrality. The first is that in country centrality,
network nodes are countries, and each country is comprised of many
prefixes.  Therefore, the paths between any two nodes in our graph is
actually the collection of paths between each pair of prefixes between
the source and destination countries. Second, the path between a pair
of prefixes is not the shortest path, but instead the country level
path of the best AS-path inferred using the techniques found in
Section \ref{sec:Populating-the-Model}. The final difference is that
prefixes can be of varying size. A prefix 12.0.0.0/8 has $2^{24}$ IP
addresses while 192.168.0.0/16 has $2^{16}$ IP addresses. Since we
assume that each country has unit traffic, we then assume that each
prefix in a country sends and receives traffic proportional to its
fraction of the country's total IP address space.  

We address the above differences with the Country Centrality
metric. We changed the $\sigma$ function to work on the best inferred
path between prefixes instead of shortest path between vertices.  We
also changed the betweenness algorithm to sum over all of the prefixes
for each country, and weight each path according to its prefix size.
The CC value of a country $\upsilon$ can be determined with the
following formula:

\[
CC(\upsilon)=\sum_{{s\neq\upsilon\neq t\in V\atop s\neq t}}\sum_{{\rho_{s\in P_{s}}\atop \rho_{t}\in P_{t}}}\left(W_{\rho_{s}}W_{\rho_{t}}\right)\sigma_{\rho_{s},\rho_{t}}(\upsilon)\]

where $\upsilon$ is a country, $P_{s}$ is the set of prefixes for
country $s$, and $W_{\rho_{s}}$ is equal to $\rho_{s}$ 's fraction
of country $s$'s prefix space $\frac{|\rho_{s}|}{\sum_{p_{i}\in P_{s}}|\rho_{i}|}$
. Here, the function $\sigma_{\rho_{s},\rho_{t}}(\upsilon)$ equals
the number of best paths between $\rho_{s}$ and $\rho_{t}$ that
transit country Since there is only one best country path between
each pair of prefixes in this function, $\sigma$ is either 1 or 0.
If each country had a single prefix, then the CC value of $\upsilon$
would be the number of shortest paths that transit $\upsilon$, which
represents the number of country-pairs that transit $\upsilon$ to
communicate. Since countries have many prefixes, and traffic between
prefixes is proportional to prefix size, a country's CC value represents
the total amount of traffic that it transits, given the stated assumptions.

To simplify CC values, we present them in this paper as normalized
values from $[0,1]$ by dividing it by the sum of traffic (with end-points
other than the country itself) that it does not transit. Therefore,
a value of one is the theoretical maximum value, suggesting that the
country transits all traffic for every country pair. Similarly, a
value of zero suggests that the country has no influence on reachability.

\subsection{Strong CC}

The CC metric estimates reachability influence based upon the best
path between each pair of prefixes. BGP routers typically have multiple
available routes to select from for each destination. Therefore, it
is possible that a country in the best path could be avoided by using
an alternate path. A network operator might intentionally try to avoid
routing through a particular country, because it is known to filter
or wiretap their data. In this subsection, we try to understand how
central countries are when alternative routes are considered.

We consider a country to be strongly  between a source and destination
prefix if all of the source's available paths include the country.
Once a router selects an alternate path, that change is propagated
throughout the network, potentially changing the tables of thousands
of other routers. Rather than attempt to measure all of the possible
network states when alternate routes are selected, we look at a snapshot
of the network's state, and determine how hard it is to avoid a country
given each router's currently available paths. The resulting measure
is called the strong country centrality SCC (SCC) metric.

\[
SCC(\upsilon)=\sum_{{s\neq\upsilon\neq t\in V\atop s\neq t}}\sum_{{\rho_{s\in P_{s}}\atop \rho_{t}\in P_{t}}}\left(W_{\rho_{s}}W_{\rho_{t}}\right)\tau_{\rho_{s},\rho_{t}}(\upsilon)\]

\comment{
\[
SCC(\upsilon)=\sum_{{s\neq\upsilon\neq t\in V\atop s\neq t}}\sum_{{\rho_{s\in P_{s}}\atop \rho_{t}\in P_{t}}}\left(W_{\rho_{s}}W_{\rho_{t}}\right)\begin{cases}
{1\atop 0} & {\frac{\sigma_{\rho_{s},\rho_{t}}(\upsilon)}{\sigma_{\rho_{s},\rho_{t}}}=1\atop otherwise}\end{cases}\]
}

In the SCC measure, $\tau_{\rho_{s},\rho_{t}}(\upsilon)$ is 1
(strongly central) when all all available paths from from $\rho_{s}$
to $\rho_{t}$ include $\upsilon$, otherwise it is 0.  Once normalized,
a value of one suggests that a country is completely unavoidable for
all paths of all country-pairs. A SCC value should be strictly less
than or equal to the same country's CC.

\section{Country Centrality Results}\label{sec:Results}
In this section we quantify the influence that countries have on
Internet reachability. We begin by determining country centrality (CC)
values from the incomplete view we have from the raw traceroute and
BGP paths described in Section~\ref{sec:Populating-the-Model}.  Then,
we test our algorithm for mapping prefix pairs to country-paths by
using the same prefixes seen in the traceroute set, but with the
inferred country-paths that provide a more complete view of the
Internet topology. This experiment shows that our metrics are robust
to the error introduced in the paths. Finally, we infer country-paths
between all pairs of prefixes and report on the CC and SCC values for
the highest-ranked countries and countries known for pervasive
censorship.

\subsection{Analysis on Directly-Observed Paths}
To start our analysis, we focus on statistics computed directly from
the paths observed in the raw traceroute and BGP data.  On the plus
side, these paths are directly observed by some source, reducing the
possibility of inference errors.  On the negative side, these data
sets provide only a partial (and potentially biased) view of paths
through the Internet, depending on the locations of iPlane monitors
(mostly PlanetLab nodes) and the vantage points where
publicly-available BGP feeds are collected.  In addition, these raw
data sets do not provide information about alternate paths, precluding
us from computing the Strong CC (SCC) metric.

\begin{table}
\begin{center}
\begin{tabular}{|l|l|l|}\cline{2-3}
\multicolumn{1}{c}{}
& \multicolumn{1}{|c|}{\textbf{TR}}
& \multicolumn{1}{|c|}{\textbf{BGP}} \\ \hline
United States & 0.335762 (1)  & 0.349493 (1) \\
Great Britain & 0.240520 (2)  & 0.187967 (2) \\
Germany       & 0.149530 (3)  & 0.165787 (3) \\
Netherlands   & 0.079117 (4)  & 0.070454 (4) \\
France        & 0.059566 (5)  & 0.061420 (5) \\
Sweden        & 0.049587 (6)  & 0.013672 (15) \\
Hungary       & 0.042618 (7)  & 0.036281 (7) \\
China         & 0.033759 (8)  & 0.045443 (6) \\
Canada        & 0.033422 (9)  & 0.034070 (8) \\  Italy         & 0.032357 (10) & 0.025297 (10) \\
Japan         & 0.024164 (11) & 0.016592 (14) \\
Denmark       & 0.022172 (12) & 0.165787 (21) \\
Russia        & 0.019994 (13) & 0.023872 (11) \\
Singapore     & 0.017008 (14) & 0.032938 (9) \\
Spain         & 0.016551 (15) & 0.013413 (16) \\
Austria       & 0.016277 (16) & 0.011704 (17) \\
South Africa  & 0.014977 (17) & 0.002211 (20) \\
Australia     & 0.010235 (18) & 0.007424 (12) \\
Serbia        & 0.007689 (19) & 0.007488 (19) \\
Norway        & 0.006837 (20) & 0.006769 (22) \\ \hline
\end{tabular}
\end{center}
\caption{Country Centrality (CC) computed directly from traceroute (TR) and BGP paths}
\label{tab:results-raw}
\end{table}

Computing the CC value of the traceroute data set was
straight-forward---we simply converted the traceroutes into
country-paths using the method described in
Section~\ref{sub:IP-Path-to-Country-Path}, and fed those paths into
the algorithm for computing the CC metric.  The results for the top 20
countries are listed in the ``TR'' column of
Table~\ref{tab:results-raw}.  Similarly, for the BGP data, we inferred
country-paths for each of the AS paths in the routing-table dumps
described in Section
\ref{sec:Populating-the-Model}.  These results are listed in the
``BGP'' column of Table~\ref{tab:results-raw}.  (Notice that the sum
of the CC values can be greater than one since multiple countries can
lie on the same path.)
The top five countries are the same in both data sets; the remaining
15 countries in the table are mostly the same, though slightly
rearranged as one might expect given the relatively small differences
in values across these countries.

The results show that three countries---the United States, Great
Britain, and Germany---have very high CC values, while many of the
commonly mentioned countries that employ censorship (e.g., China and
Iran) have relatively little influence over global reachability.
European countries are heavily represented in the table, including
some countries with higher rankings than we expected---such as the
Netherlands, Sweden, and Hungary.  We suspect that the relatively
large number of (small) countries in Europe cause a large number of
European countries to rely on other countries in the same region for
connectivity to the rest of the Internet.  In addition, these results
may be, at least in part, an artifact of the incomplete perspective of
the raw traceroute and BGP data; as seen in the next section, these
three countries drop somewhat (though admittedly not dramatically) in
the ranking when we use the more complete, inferred paths.

\subsection{Validation of Inference of Country Paths}
The CC results from the raw traceroute and BGP data, while
interesting, represent only a tiny sample of the Internet's
country-paths.  Still, these data sets are useful for
validating our country-path inference technique.
The validation experiment compares the CC results of real country-paths
(directly mapped IP addresses to countries) to inferred country-paths
(country-paths inferred from only the source and destination IP addresses).
The inference algorithm was trained on the training sets of traceroutes
and BGP RIBs. Then, we used the primed country-path inference algorithm
to infer paths between the (source,destination) IP address pairs in
the testing traceroute set. It is possible that the testing traceroute
may have a source IP from an AS in the RIB training set. The algorithm
would then have a known AS-path to inference, which would invalidate
the experiment. To prevent such overlap from affecting our results,
we ignored such traceroutes in the experiment.

\comment{
\begin{figure}
\begin{centering}
 \includegraphics[width=1.05\columnwidth]{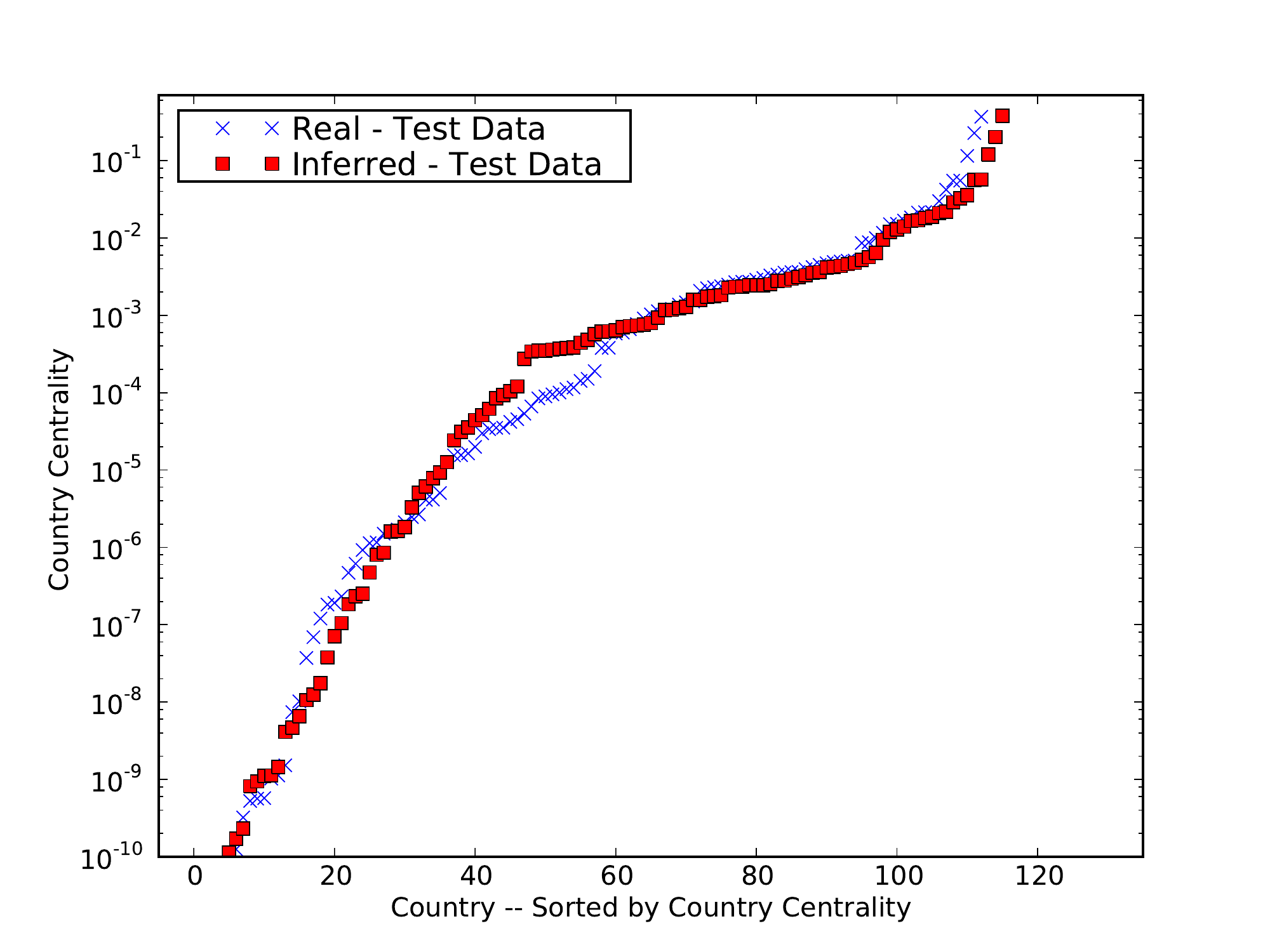}
\par\end{centering}
\caption{Actual versus Predicted Country Centrality.
Country Centrality (CC) is plotted on the log-scaled y-axis for
countries arranged on the x-axis in rank order of their actual CC
value.  Xs show the actual CC values computed from the
traceroute data set. Squares represent CC values that were inferred
using the Country Path Algorithm.}
\label{fig:CC-Validation}
\end{figure}
}
\begin{figure}
\begin{centering}
 \includegraphics[width=1.05\columnwidth]{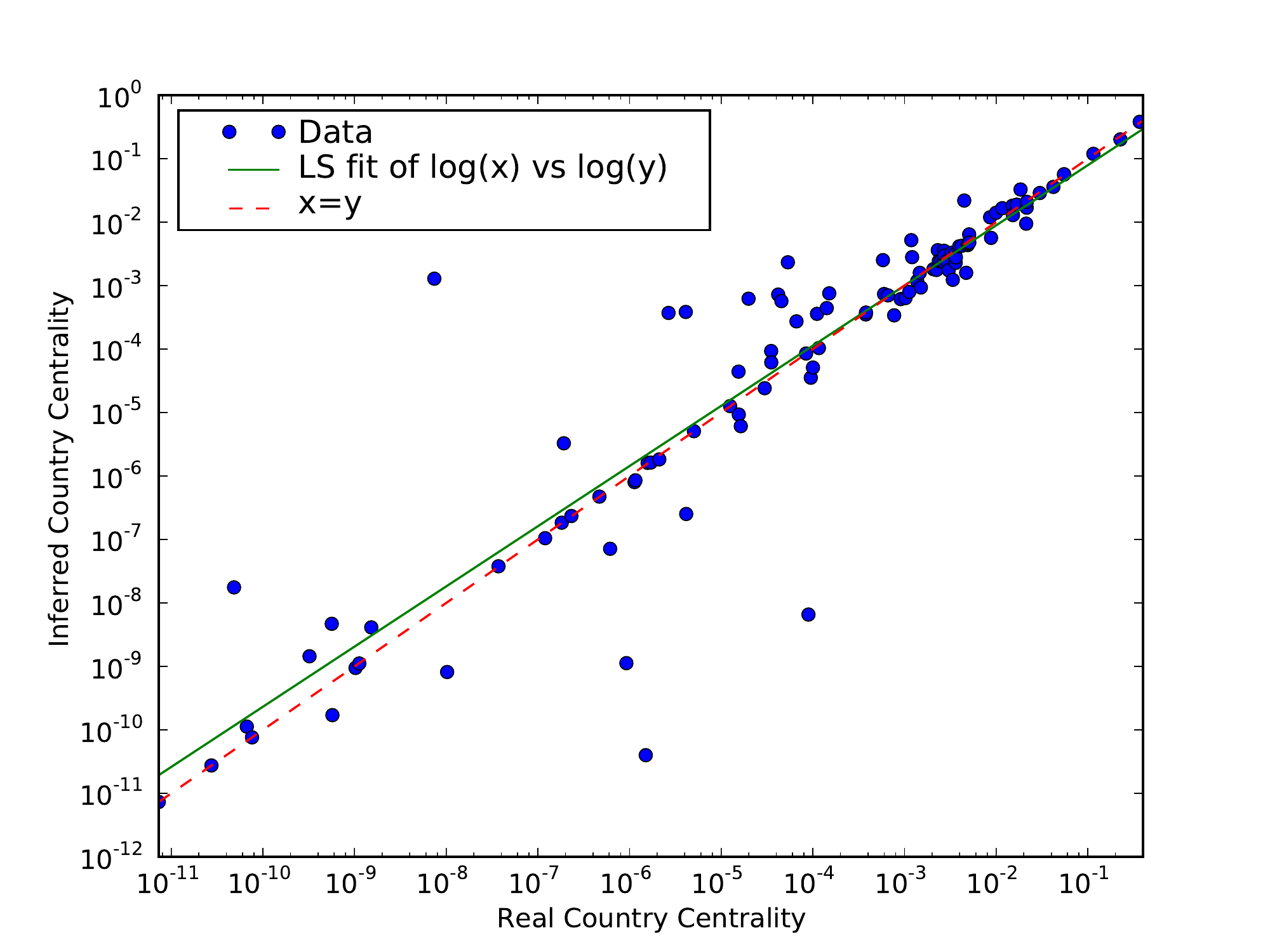}
\par\end{centering}
\caption{Actual versus Predicted Country Centrality.  Predicted
  Country Centrality (CC) (log-scaled y-axis) is plotted against the
  actual CC for the same countries (log-scaled x-axis). Because there
  are so many small values, the data is fit in log(y) vs log(x) space
  to prevent overfitting the large values.  The least squares linear
  fit is a solid line and the ideal $x=y$ line is dashed.}
\label{fig:CC-Validation-Scatter}
\end{figure}

We plot the results of the inferred paths against what are believed to
be accurately inferred {}``real'' country-paths in
Figure~\ref{fig:CC-Validation-Scatter}.  Both axis are log scaled to
show the countries with low centrality in greater detail.  Ideally,
the data points would reside along the dotted $x=y$ line, suggesting
that the CC of the real paths and inferred paths are the same.  Many
of them, especially the larger values, do lie closely along that line.
Only a few extreme outliers exist, and they have relatively low CC
values.  We produced a least squares linear fit of $log(x)$ vs
$log(y)$.  It is plotted as a solid line, and has slope 0.94, with an
$R^2$ of 0.84. This experiment leads us to believe that while there is
inference error, the CC measurement is robust enough to the noise that
the resulting values are meaningful.

\subsection{Analysis on More Complete Country Paths}
Because our inferred results match the CC values of the real paths so
well, we inferred the entire set of country paths between all 290,682
routable prefixes found in our collection of RIBs. The country-path
inference algorithm was trained on the full traceroute and RIB data
sets. In total, the entire computation took two days to run when
spread over 14 processors.  %
Figure~\ref{fig:CC-Global} plots the CC values of all countries,
sorted by their CC values.  Not surprisingly, the vast majority of
countries have very small CC values.  We list the top 20 countries in
the ranking in the ``CC'' column in Table~\ref{tab:results-inf}.  The
list of countries has a significant overlap with
Table~\ref{tab:results-raw}.  The top five countries are the same,
with just France (\#4) and the Netherlands (\#5) swapped in ranking
between the two lists.

\begin{figure}
\begin{centering}
\includegraphics[width=1.05\columnwidth]{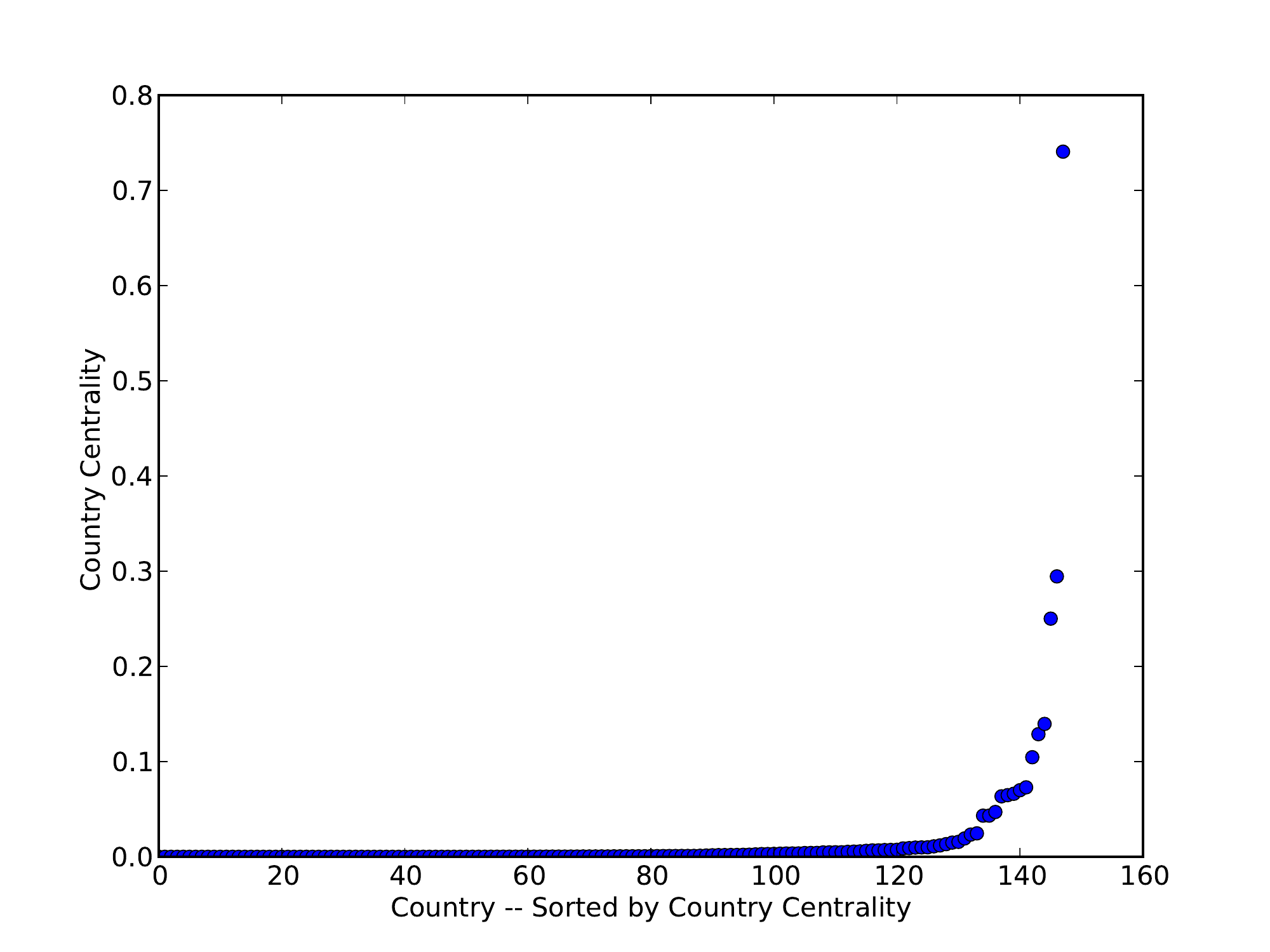}
\par\end{centering}
\caption{Country Centrality (CC) on more complete, inferred country-paths.
Countries are displayed on the x-axis, sorted by their CC values, and
CC values are displayed on the y-axis.}
\label{fig:CC-Global}
\end{figure}

\begin{table}
\begin{center}
\begin{tabular}{|l|l|l|}\cline{2-3}
\multicolumn{1}{c}{}
& \multicolumn{1}{|c|}{\textbf{CC}}
& \multicolumn{1}{|c|}{\textbf{SCC}} \\ \hline

United States  & 0.740695 (1) & 0.546789 (1) \\
Great Britain  & 0.294532 (2) & 0.174171 (2) \\
Germany        & 0.250166 (3) & 0.124409 (3) \\
France         & 0.139579 (4) & 0.071325 (4) \\
Netherlands    & 0.128784 (5) & 0.051139 (5) \\
Canada         & 0.104595 (6) & 0.045357 (6) \\
Japan          & 0.072961 (7) & 0.027095 (11) \\
China          & 0.069947 (8) & 0.030595 (10) \\
Australia      & 0.066219 (9) & 0.037885 (8) \\
Hungary        & 0.064767 (10) & 0.023094 (14) \\
Singapore      & 0.063522 (11) & 0.043445 (7) \\
Italy          & 0.047068 (12) & 0.027088 (12) \\
Spain          & 0.043248 (13) & 0.025370 (13) \\
Russia         & 0.043228 (14) & 0.035191 (9) \\
Austria        & 0.024632 (15) & 0.010501 (17) \\
Sweden         & 0.023350 (16) & 0.009785 (19) \\
South Africa   & 0.019294 (17) & 0.013778 (15) \\
Denmark        & 0.015684 (18) & 0.008101 (21) \\
Serbia         & 0.014935 (19) & 0.012312 (16) \\
Switzerland    & 0.013302 (20) & 0.003865 (35) \\ \hline
\end{tabular}
\end{center}
\caption{Country Centrality (CC) and Strong Country Centrality (SCC) computed using inferred country paths}
\label{tab:results-inf}
\end{table}

\begin{figure}
\begin{centering}
\includegraphics[width=1.05\columnwidth]{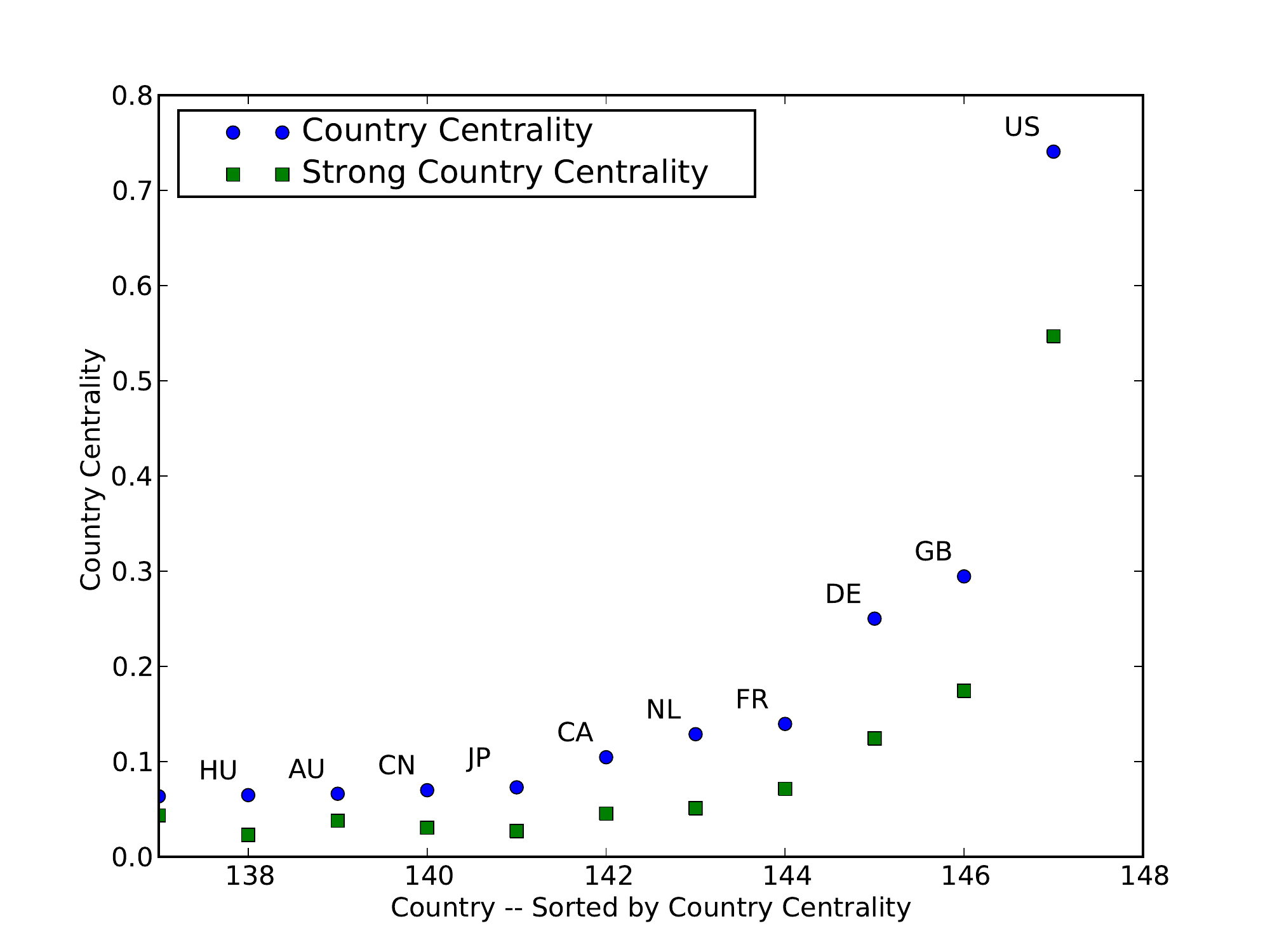}
\par\end{centering}
\caption{Strong CC (Zoomed). The top 10 countries (in terms of CC value)
are displayed on the x-axis, sorted by their CC values. The CC values
are displayed on the y-axis. The squares represent the Strong CC values
of each respective country and have the same scale as the CC data.}
\label{fig:OA-Global-Zoom}
\end{figure}

Surprisingly, the U.S. has a significantly higher CC value in
Table~\ref{tab:results-inf}---nearly \emph{double} the CC value in
Table~\ref{tab:results-raw}.  We suspect that this is caused by the
sampling bias in the traceroute and BGP data sets. For instance, the
incomplete data sets likely over sample the routes from countries that
have well-distributed connections to the Internet (such as European
countries) and under sample countries with less rich connectivity (such
as those in South America) that often rely on the United States for
reachability to the rest of the Internet.  This disparity points out
the importance of having a more complete view of
country-paths---possible because of the inference algorithms we used
to compute paths from vantage points that do not run iPlane monitors
or provide BGP measurement feeds.

Next, we investigate the Strong CC (SCC) of each country. This is an
estimate of the difficulty in circumventing a given country, even if
alternate routes are used. The results are shown in the ``SCC'' column
of Table~\ref{tab:results-inf}.  The table shows that the top three
countries have high SCC values, suggesting that they are hard to avoid
even using alternate paths.  We also show the top 10 CC and SCC
countries in Figure \ref{fig:OA-Global-Zoom}.  Not surprisingly, the
U.S. is especially difficult to avoid, especially for countries (e.g.,
in South America) that connect directly to the U.S. for connectivity
to the Internet.

\begin{table}
\begin{center}
\begin{tabular}{|l|c|c|} \cline{2-3}
\multicolumn{1}{c|}{} &
 \textbf{CC} &
 \textbf{SCC} \\ \hline

China        & 0.069947 (8) & 0.030595 (10)\\
Vietnam      & 0.007087 (30) & 0.003916 (34)\\
South Korea  & 0.003548 (44) & 0.001044 (54)\\
Saudi Arabia & 0.003286 (47) & 0.001722 (49)\\
U.A.E.       & 0.000839 (65) & 0.000541 (63)\\
Pakistan     & 0.000274 (81) & 0.000265 (74)\\
Iran         & 1.12e-05 (105) & 9.48e-06 (101)\\
Yemen        & 1.06e-07 (131) & 7.50e-08 (130)\\
Oman         & 2.64e-08 (138) & 2.64e-08 (133)\\
Myanmar      & 0 & 0\\
North Korea  & 0 & 0\\
Sudan        & 0 & 0\\
Syria        & 0 & 0\\
\hline
\end{tabular}
\end{center}
\caption{CC and SCC values of countries with pervasive censorship.
Countries with 0 values were not found to transit \emph{any} international
traffic.}
\label{tab:Pervasive}
\end{table}

Finally, we consider the countries that are known for significant
censorship.  When Internet censorship is discussed, China, Iran, Saudi
Arabia, and Pakistan are commonly mentioned as countries that filter
Internet traffic.  According to the OpenNet Initiative~\cite{ONI},
these four countries along with eight others partake in pervasive
traffic filtering. The CC values of each of these countries is shown
in Table~\ref{tab:Pervasive}. Aside from China (with a CC of 0.07),
these countries appear to have very little influence over global
reachability.  We were initially surprised to see that South Korea has
a relatively low CC value ($0.004$), given the significant penetration
of the Internet in the country.  However, the large deployments of
broadband connectivity for end users need not relate to whether Korean
ISPs play an important role in transit service for other countries.

\comment{ 
In this section we measure the influence that countries have on Internet
reachability. We begin by determining country CC values from the
partial views of the traceroutes and BGP RIBs described in Section
\ref{sec:Populating-the-Model}. Then, we test our prefix-pair to
country-path algorithm by running the measurements on the same prefixes
as found in the traceroute set, but with inferred country-paths. This
experiment shows that our metrics are robust to the error introduced
in the paths. Finally, we infer country-paths between each pair of
prefixes and measure the global CC and SCC of each country.

\subsection{CC of Traceroutes and RIBS}

\begin{figure}
\begin{centering}
\includegraphics[width=1.05\columnwidth]{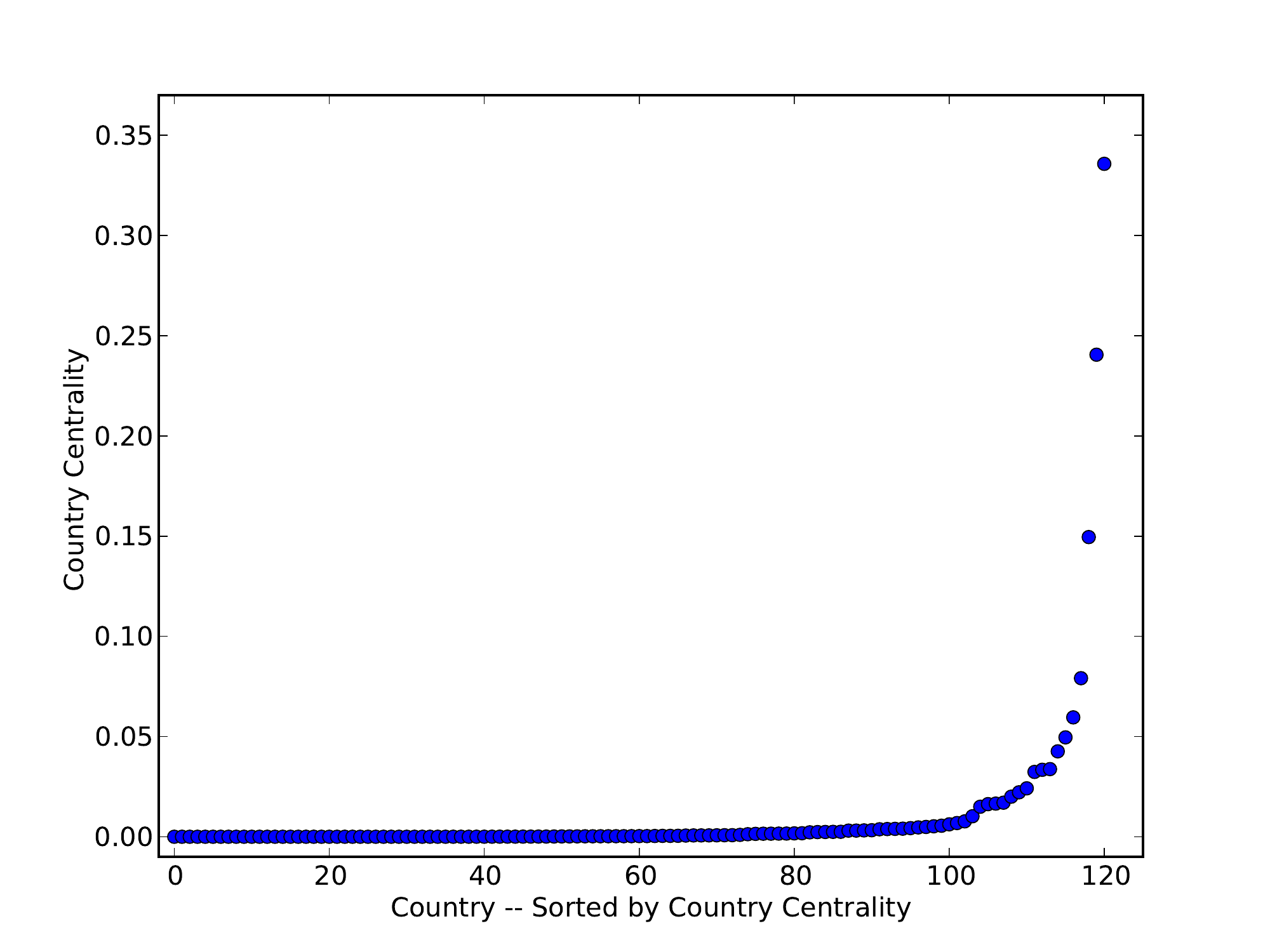}
\par\end{centering}

\caption{CC of the traceroute data set. Incomplete paths were not included
(due to lack of response to traceroute or unknown country for an IP).
The x-axis lists the countries, in order of CC from least to greatest.
The y-axis shows the countries CC. Notice that the sum of the CC
values can be greater than one since multiple countries can exist
on the same path. }
\label{fig:CC-Trace}
\end{figure}
\begin{figure}
\begin{centering}
\includegraphics[width=1.05\columnwidth]{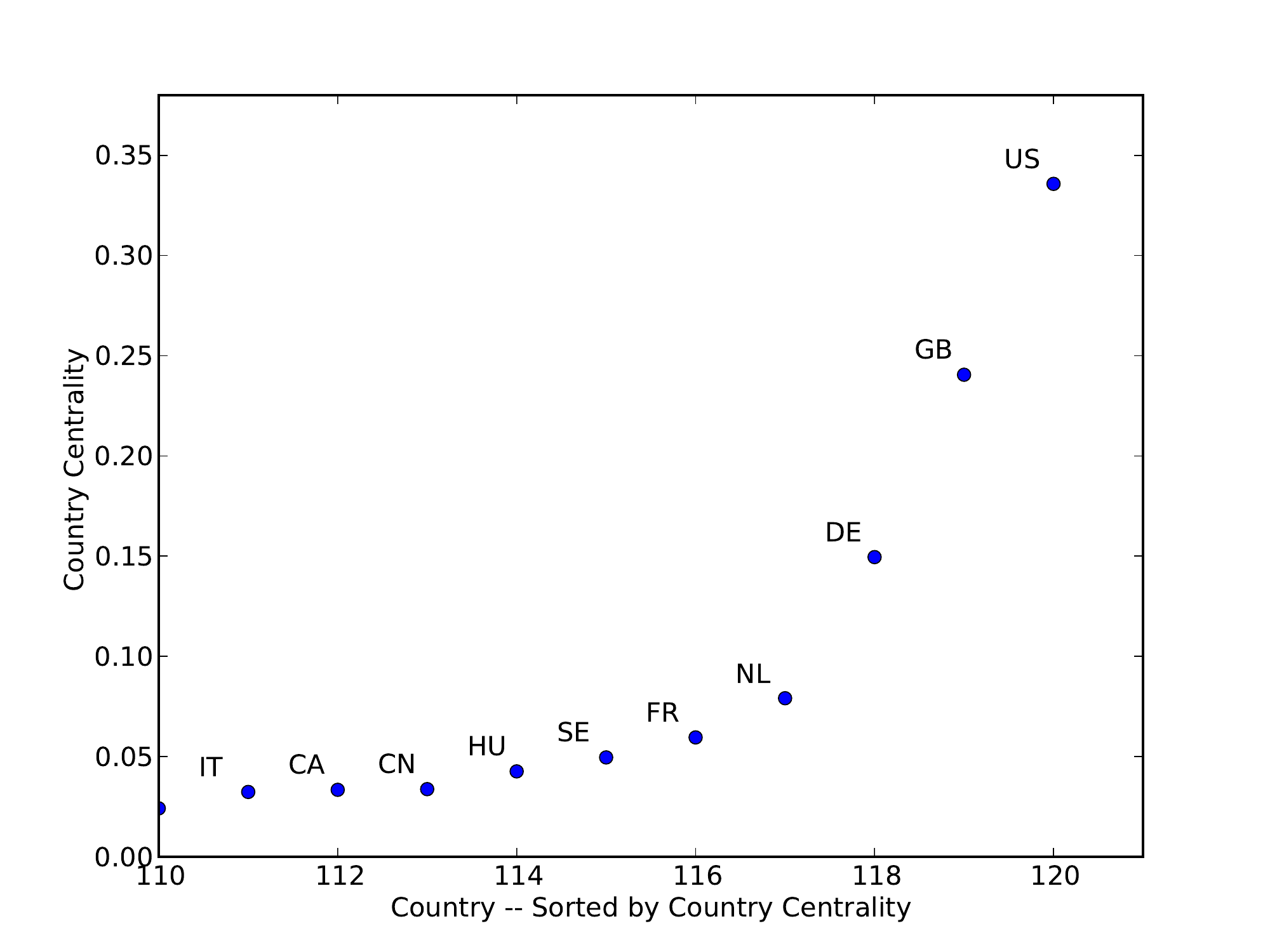}
\par\end{centering}

\caption{This figure zooms into the highest 10 values of Figure
  \ref{fig:CC-Trace} and labels the individual countries.  The two
  letter country codes are from ISO 3166.}
\label{fig:CC-TraceZoom}
\end{figure}

\begin{figure}
\begin{centering}
\includegraphics[width=1.05\columnwidth]{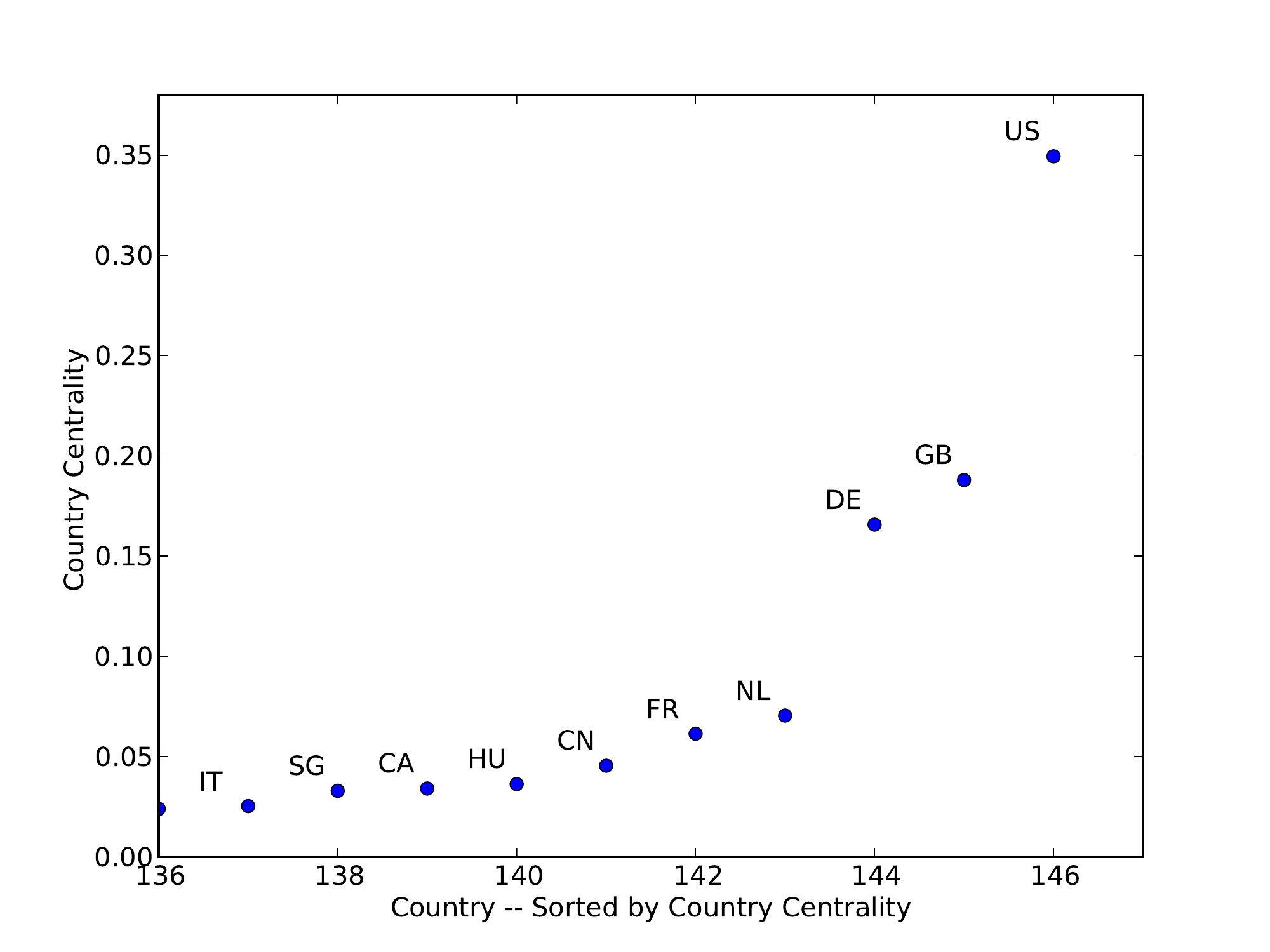}
\par\end{centering}

\caption{This figure is like Figure \ref{fig:CC-Trace} but it shows CC values
inferred from known AS-paths. Note that the top 5 countries are the
same as in Figure \ref{fig:CC-Trace} and the remaining countries
are mostly the same, but slightly rearranged (which is to be expected
as the difference in CC values decreases).}
\label{fig:CC-BGPZoom}
\end{figure}

Measuring the CC value of the traceroute data set is straight-forward.
We simply converted the traceroutes into country-paths using the method
described in Section \ref{sub:IP-Path-to-Country-Path}, and fed those
paths into the CC metric. The resulting CC values are shown in Figures
\ref{fig:CC-Trace} and the top ten countries are displayed more
closely in \ref{fig:CC-TraceZoom}. The results show us that three
countries (the United States, Great Britain, and Germany) have very
high CC values and also suggest that many of the commonly mentioned
countries that employ censorship (e.g. China, and Iran) have relatively
little influence over global reachability. 

Next, we inferred country-paths for each of the AS-paths in the RIBs
described in Section \ref{sec:Populating-the-Model}, and plotted
the CC results in Figure \ref{fig:CC-BGPZoom}. The results of this
experiment are very close to the traceroute results, even though the
data sets are collected from separate observation points and this
experiment involved inferenced country-paths for each AS path.

\subsection{Validation}

\begin{figure}
\begin{centering}
  \includegraphics[width=1.05\columnwidth]{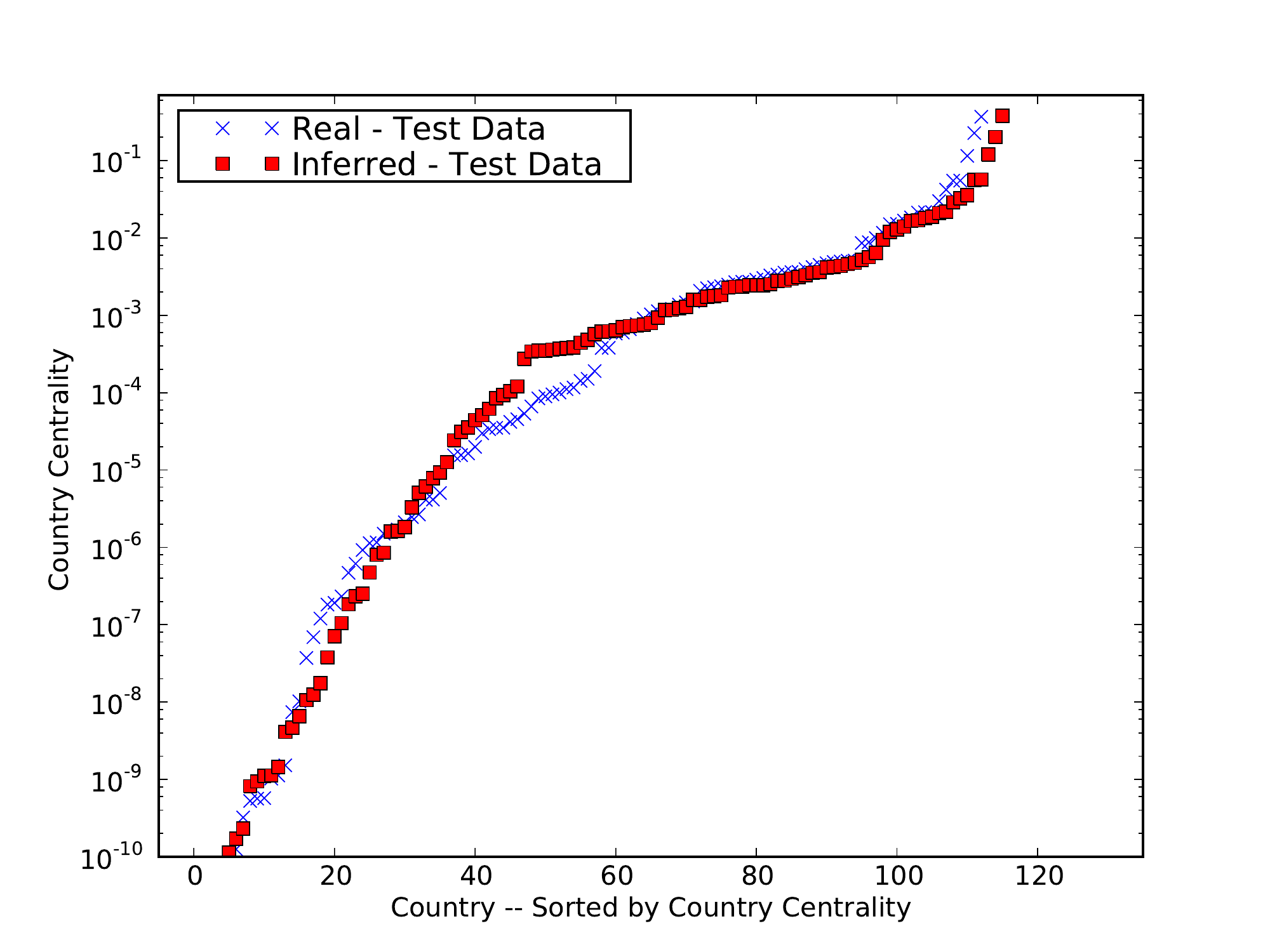}
\par\end{centering}

\caption{Validation of country centrality. The y-axis is log scaled to
  show greater definition. The countries are displayed on the x-axis,
  sorted by their CC values. The CC values are displayed on the
  y-axis. The circles represent the CC values derived from the
  traceroute testing data set. The squares represent the CC values
  derived from inferred country level paths between each pair of
  (source,destination) addresses in the testing set.}
\label{fig:CC-Validation}
\end{figure}

The CC results from the BGP and Traceroute data are interesting,
but they represent only a tiny sample of the Internet's country-paths.
Further, those data sets contain only the best paths, and SCC values
cannot be determined from them. With our country-path inference
technique, it would be possible to analyze every country-path (and
alternate paths). In this sub-section we determine if the country-path
inference technique described earlier is accurate enough to extract
meaningful CC and SCC values from. 

The validation experiment compares the CC results of real country-paths
(directly mapped IP addresses to countries) to inferred country-paths
(country-paths inferred from only the source and destination IP addresses).
The inference algorithm was trained on the training sets of traceroutes
and BGP RIBs. Then, we used the primed country-path inference algorithm
to infer paths between the (source,destination) IP address pairs in
the testing traceroute set. It is possible that the testing traceroute
may have a source IP from an AS in the RIB training set. The algorithm
would then have a known AS-path to inference, which would invalidate
the experiment. To prevent such overlap from affecting our results,
we ignored such traceroutes in the experiment.

We plot the results of the inferred paths against what are believed
to be accurately inferred {}``real'' country-paths in Figure \ref{fig:CC-Validation}.
The y-axis is log-scaled so that the two plots can be more easily
compared across all scales. One important point to notice is that
there are more countries in the inferred data set. This is because
some of the inferred paths contained countries that did not exist
in the testing traceroutes. This is likely due to the fact that the
algorithm is primed on all of the testing set of routes, while only
complete traceroutes (no time-out values and all IP addresses can
be converted to countries) from the testing set are used to inference.
Clearly, there is significant agreement between the two plots, even
down to the $10^{-10}$ scale. This experiment leads us to believe
that while there is inference error, the CC measurement is robust
enough to the noise that the resulting values are meaningful.

\subsection{Global CC and SCC}

\begin{table}
\begin{centering}
{\tiny }\begin{tabular}{|c|c|}
\hline 
{\tiny Country} & {\tiny CC Value}\tabularnewline
\hline
\hline 
{\tiny China} & {\tiny 0.0537}\tabularnewline
\hline 
{\tiny Vietnam} & {\tiny 0.0056}\tabularnewline
\hline 
{\tiny Saudi Arabia} & {\tiny 0.0031}\tabularnewline
\hline 
{\tiny South Korea} & {\tiny 0.0017}\tabularnewline
\hline 
{\tiny United Arab Emirates} & {\tiny 0.0008}\tabularnewline
\hline 
{\tiny Pakistan} & {\tiny 0.0003}\tabularnewline
\hline 
{\tiny Iran} & {\tiny 1.07e-05}\tabularnewline
\hline 
{\tiny Yemen} & {\tiny 6.918e-08}\tabularnewline
\hline 
{\tiny Oman} & {\tiny 2.4395e-08}\tabularnewline
\hline 
{\tiny Myanmar} & {\tiny nil}\tabularnewline
\hline 
{\tiny North Korea} & {\tiny nil}\tabularnewline
\hline 
{\tiny Sudan} & {\tiny nil}\tabularnewline
\hline 
{\tiny Syria} & {\tiny nil}\tabularnewline
\hline
\end{tabular}
\par\end{centering}{\tiny \par}
\caption{CC values of countries known to have censorship policies. Countries will
  nil entries did not transit traffic in any international
  paths in our data set.}
\label{tab:Pervasive}
\end{table}
\begin{figure}
\begin{centering}
\includegraphics[width=1.05\columnwidth]{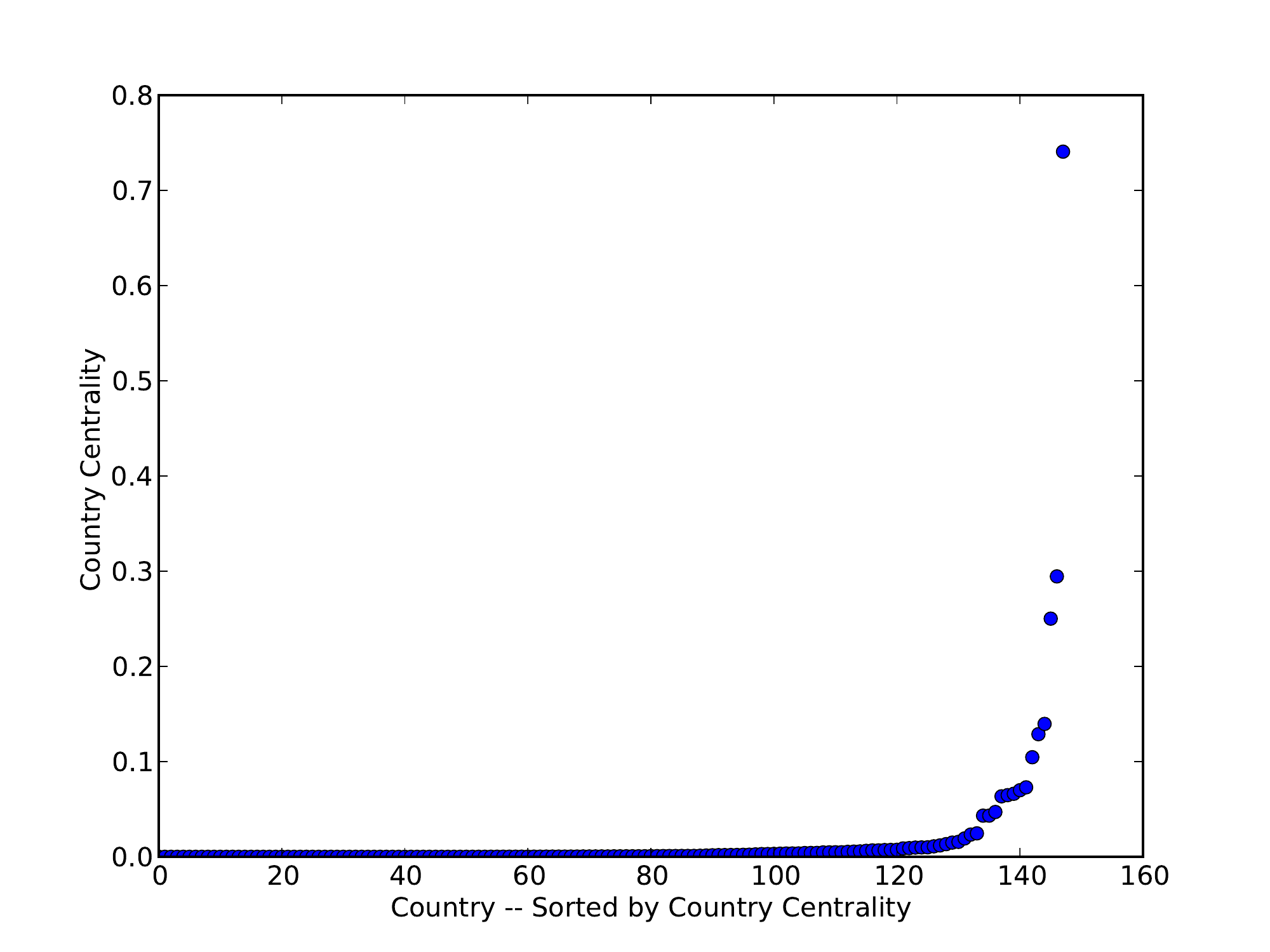}
\par\end{centering}

\caption{Global country centrality. The countries are displayed
on the x-axis, sorted by their CC values. The CC values are displayed
on the y-axis. }
\label{fig:CC-Global}
\end{figure}
\begin{figure}
\begin{centering}
\includegraphics[width=1.05\columnwidth]{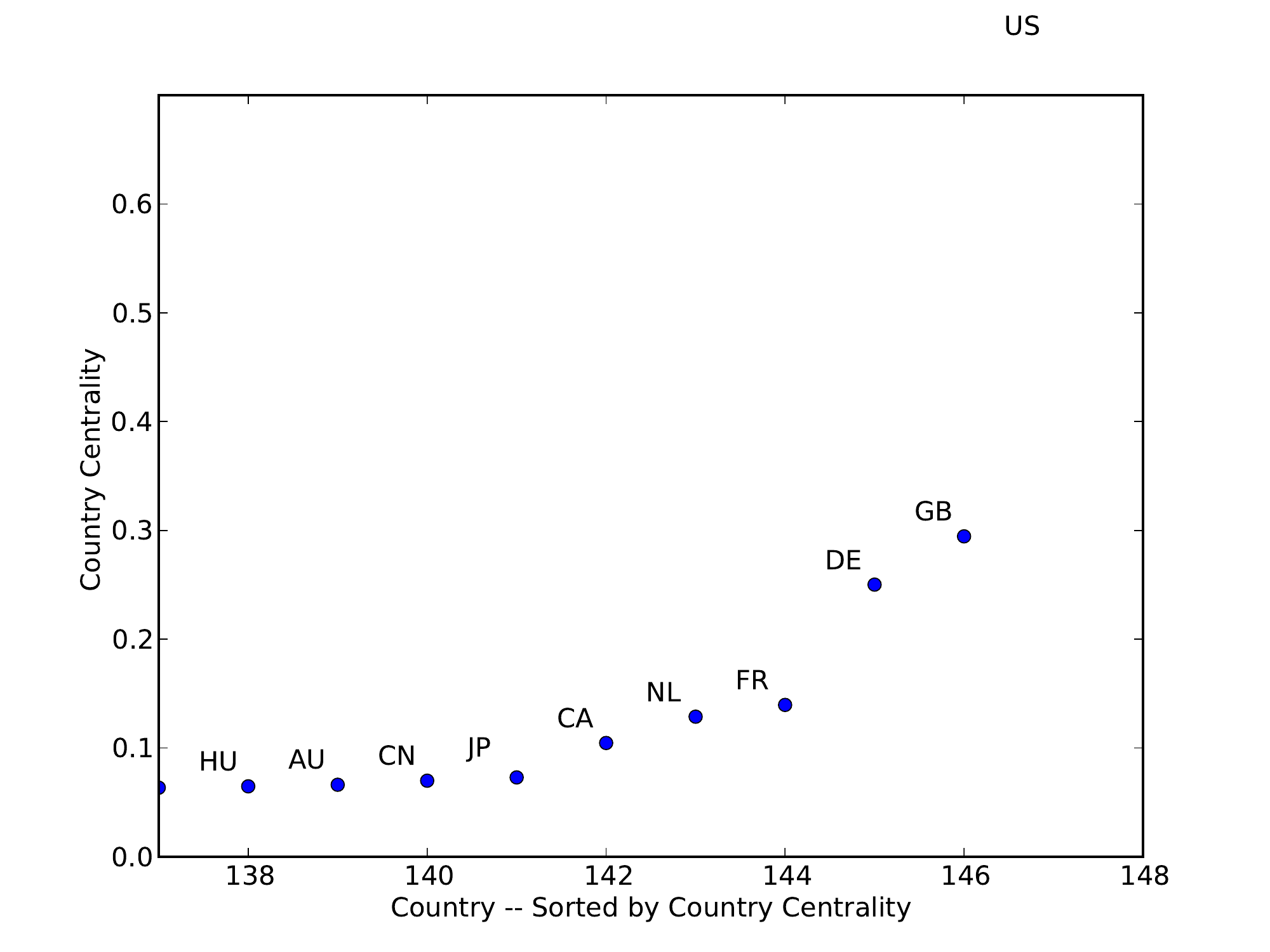}
\par\end{centering}

\caption{Global country centrality (zoomed in). The top 10
countries (in terms of CC value) are displayed on the x-axis, sorted
by their CC values. The CC values are displayed on the y-axis. }
\label{fig:CC-Global-Zoom}
\end{figure}
\begin{figure}
\begin{centering}
\includegraphics[width=1.05\columnwidth]{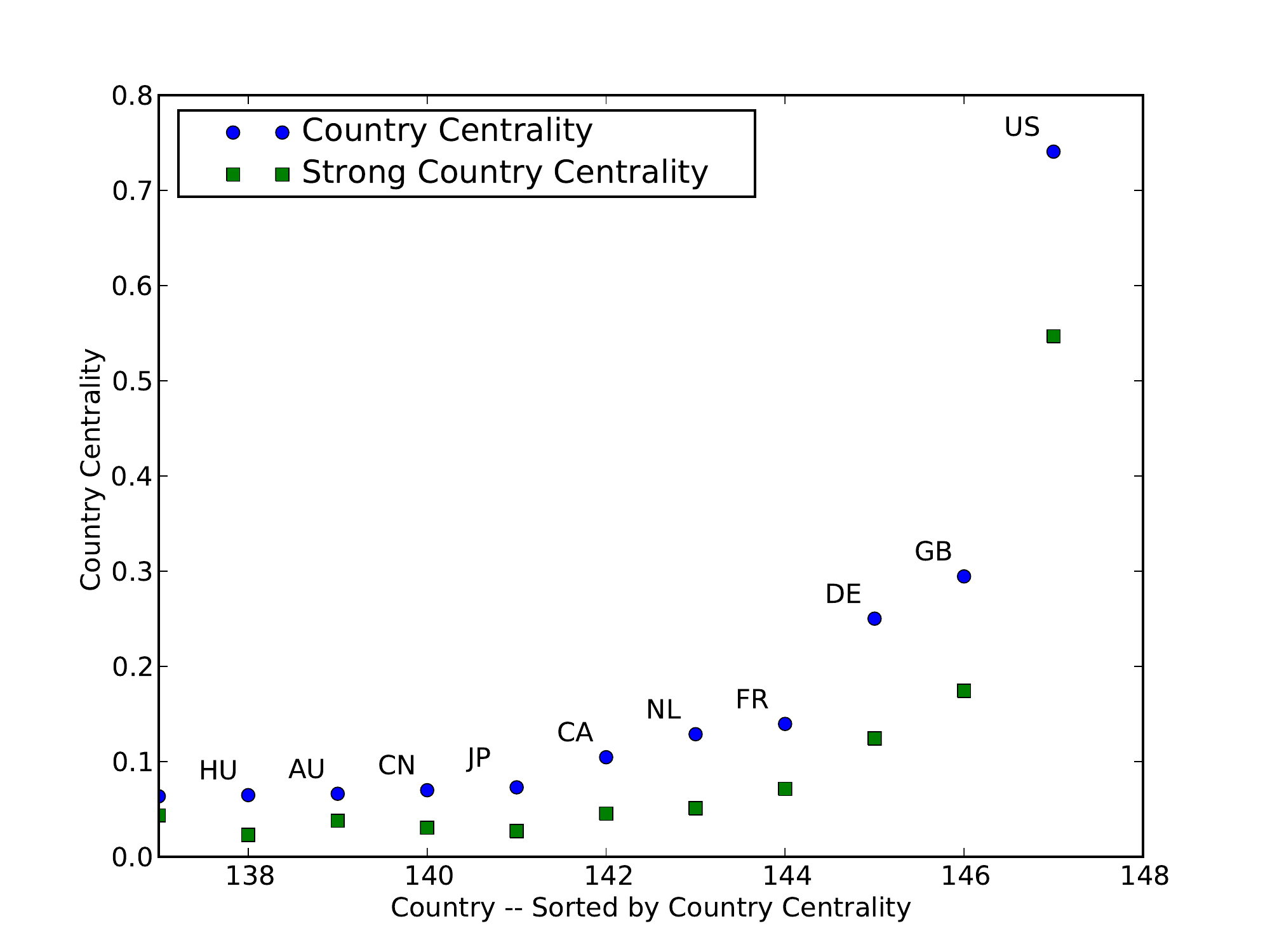}
\par\end{centering}

\caption{Strong CC (Zoomed). The top 10 countries (in terms of CC value)
are displayed on the x-axis, sorted by their CC values. The CC values
are displayed on the y-axis. The squares represent the SCC values
of each respective country. We use the same y-axis for both plots
as the CC and SCC values have similar meaning.}
\label{fig:OA-Global-Zoom}
\end{figure}

Because our inferred results match the CC values of the real paths
so well, we inferred the entire set of country paths between all 290,682
routable prefixes found in our collection of RIBs. The country-path
inference algorithm was trained on the full traceroute and RIB data
sets. In total, the entire calculation took two days to compute and
was spread over 14 cpus. 

We first present the global CC of each country. This is shown in
Figure \ref{fig:CC-Global} and the top ten countries are shown in
focus in Figure \ref{fig:CC-Global-Zoom}. These values show significant
agreement with the earlier results derived from small data sets. Nine
out of the top 10 countries exist in the results derived from known
BGP paths and the global results. Eight out of the top 10 countries
exist in both the global results and those derived from known traceroutes.
Most of the countries have stable CC values between all of the experiments
in this section. Surprisingly, the United States nearly doubles its
CC value at the global scale. We suspect that this is due to bias
in the traceroute and BGP data sets. For instance, the data sets may
over sample routes from countries that have well distributed connections
to the Internet (such as European countries) and under sample countries
with centralized connections (such as those in South America) that
often rely on the United States for routing. 

When Internet censorship is discussed, countries such as China, Iran,
Saudi Arabia, and Pakistan are commonly mentioned as examples of countries
that filter traffic. According to the OpenNet Initiative ~\cite{ONI},
these countries along with eight others partake in pervasive traffic
filtering. The CC values of each of these countries is shown in Table
\ref{tab:Pervasive}. Aside from China (CC of 0.07), these countries
appear to have very little influence over global reachability. 


Finally, we measure the Strong CC (SCC) of each country. This
is an estimate of how difficult it is to avoid routing through a given
country, even if alternate routes are used. The results are shown
in Figure \ref{fig:OA-Global-Zoom}, and are plotted against the CC
values of the same countries. The figure shows that the top three
countries have high SCC values, suggesting that they are hard to
avoid even using alternate paths. 
}

\section{Discussion and Future Work}\label{sec:Discussion}
There are several potential sources of bias in the data sets we used,
which could potentially impact the results.

First, it is believed that the Internet's topology is significantly
larger than what can be observed in BGP RIBs~\cite{oliveira08}.  For
example, peer-peer connections are only visible to customers of the
peers (due to the valley-free rule) and are thus difficult to
find~\cite{Chang:04}.  Fortunately, it is believed that
customer-provider edges are well represented in the observed RIBs.
The topologies that we extracted from the RIBs support these
suppositions.  As shown in~\ref{sub:Prefix-Pair-toAS}, the number of
peer-peer edges increases by 90\% between the testing set and the
total set while customer-provider edges only increased by 5\%.
Peer-peer edges typically have less impact on routing than
customer-provider edges, since only the downstream customers of the
two peers can route through peer-peer edges.  In addition, we suspect
that peer-peer edges, for the most part, arise between ASes in the
same country, or at least the same small geographic region (e.g.,
between two countries in Europe), which would also limit their
influence on the international flow of traffic through the Internet.
Still, the missing edges could have impact on the results of our
measurements. To test this, we plan to run our algorithms on multiple
inferred and generated~\cite{chang06,holme08} topologies, including
traceroute measurements collected from larger number of vantage
points~\cite{netdimes}.

Beyond the question of bias, we would also like to study the evolution
of country centrality over time. It has been suggested that the United
States transits a smaller fraction of total traffic than in the past.
It would be interesting to know if the United States has also become
less central in terms of reachability, and if so why.  Which countries
are becoming more central over time and which less so?  It would also
be interesting to know how our results would change if we incorporated
more realistic models of interdomain traffic~\cite{feldmann:04}.
A more long-term question involves understanding
the economically-driven strategies that single countries or small groups
of countries could adopt, either to enhance their own centrality or to
reduce the centrality of other countries (e.g., such as overlay routing).
There may also be other network
measures that are of interest.  Deletion impact or measures that
incorporate some component of traffic are two obvious directions.

Finally, the paths traversed by domestic traffic would also be
interesting to study. What fraction of domestic paths (those that have
a source and destination within the same country) are actually routed
through another country?  Answering this question would provide
insight into the influence that foreign nations have over a country's
domestic routing and security, and would shed light on a question
posed in~\cite{PAA} concerning whether warrantless wiretapping on
links connecting one country to another might inadvertently capture
some purely domestic traffic.  The framework developed in this paper
could be extended to address that question.


\section{Related Work}\label{sec:Related-Work}
We are unaware of previous work measuring the impact individual
countries have on the flow of international traffic in the Internet.
However, our results rely on earlier work on network centrality,
Internet topology measurement, AS-relationship inferencing, AS-path
inferencing, and studies of Internet censorship.  In this section we
briefly review the projects most relevant to this paper.

In addition to Qiu et al.'s work~\cite{Qiu05} discussed earlier, there are at least
two other methods for inferring AS-paths that are prefix specific.
M\"uhlbauer et al.~\cite{Muhlbauer06} showed that when an AS has multiple
routers distributed across many locations, more
than one router needs to be simulated to capture all of the routing
diversity within the AS.  By simulating multiple quasi-routers per AS,
they were able to predict AS-paths with relatively high accuracy
(reported 65\%); however, the high overlap between their testing
and training data sets makes it difficult to compare the accuracy
of their technique with ours.
and more computationally efficient.  This allowed us to study all
290,000 prefixes rather than the 1000 prefixes reported in M\"uhlbauer et al.


Another AS-path inference algorithm was developed by Madhyastha et
al.,~\cite{Madhyastha06} who used a structural approach to AS-path prediction.  They
began with known traceroutes from the iPlane project and used them to
infer IP-level paths for chosen src/dest pairs.
The algorithm works by searching for the closest observation point to
the source prefix (by examining a few sample traceroutes from the
source) and then uses the known iPlane paths to infer the remaining
paths from the source.  They do not report the accuracy of the
IP-level paths, but we are interested
in investigating this technique in future work as an alternate way
to infer country paths.

Finally, there has been an enormous amount of work developing
statistical measures of network properties~\cite{Freeman77,Freeman79,Sabidussi66}, including
preferential attachment models~\cite{ba:model} and many models of the AS network~\cite{alv:inet,holme08,chang06,fkp:model,yook:inet}.  Some of this
work measures node centrality by the impact it would have on network
connectivity if the AS was deleted, known as deletion impact~\cite{chn:perc,holme08}. A parallel can potentially be made between
node deletion and censorship.  For example, deleting a country from the
network is conceptually similar to all other ASes collectively
routing around that country.


\section{Conclusions}\label{sec:Conclusion}
As government control over the treatment of Internet traffic becomes
more common, many people will want to understand how international
reachability depends on individual countries and to adopt strategies
either for enhancing or weakening the dependence on some countries.
The work presented in this paper is an initial step towards providing
the algorithms and tools that will be needed to understand and manage
nation-state routing.

In particular, we discussed the problems associated with understanding
routing patterns at the country level, which is the level at which
most censorship and wiretapping policies are mandated.  We then
described algorithms and data sources to infer country-level paths
from traceroute probes and AS-level BGP data, and we validated those
algorithms against different samples of the same kinds of data.  Next
we discussed metrics for comparing the relative importance of
different countries in current routing topologies.  Finally, we used
the algorithms to infer a country path between each pair of IPv4
prefixes and then applied the metrics to the paths to obtain initial
results.

It is not surprising that the results show the dominance of the
U.S. at the country routing level.  However, other countries appear to
have either more or less importance than one might expect.  For
example, both Great Britain and Germany are second only to the U.S.
in centrality, while Japan, China, and India are only 8th, 10th, and
32nd respectively.  Collectively, these results show that the ``West''
continues to exercise disproportionate influence over international
routing, despite the penetration of the Internet to almost every
region of the world, and the rapid development of China and India.
Beyond what the results tell us about the Internet today, we see the
methods described in this paper as helping network designers, policy
makers, and researchers better understand the likely impact of
national policies on user privacy and the access to politically
or socially sensitive content.

\bibliographystyle{ieeetr}
\bibliography{nation-state-routing}
\end{document}